\documentclass[
 preprint,
 amsmath,amssymb,
 aps,
prb,
]{revtex4-1}

\usepackage{graphicx}
\usepackage{dcolumn}
\usepackage{bm}
\usepackage{hyperref}
\usepackage[mathlines]{lineno}
\usepackage{commath}
\usepackage{physics}
\usepackage{color}

\begin{document}

\preprint{APS/123-QED}

\title{Intrinsic persistent spin helix in two-dimensional group-IV monochalcogenide $MX$ ($M$: Sn, Ge; $X$: S, Se, Te) monolayer}

\author{Moh. Adhib Ulil Absor}
\email{adib@ugm.ac.id} 
\affiliation{Department of Physics, Universitas Gadjah Mada, Sekip Utara, BLS 21 Yogyakarta Indonesia.}%

\author{Fumiyuki Ishii}%
\affiliation{Nanomaterial Research Institute, Kanazawa University, 920-1192, Kanazawa, Japan.}%

\date{\today}

\begin{abstract}
Energy-saving spintronics are believed to be implementable on the systems hosting persistent spin helix (PSH) since they support an extraordinarily long spin lifetime of carriers. However, achieving the PSH requires a unidirectional spin configuration in the momentum space, which is practically non-trivial due to the stringent conditions for fine-tuning the Rashba and Dresselhaus spin-orbit couplings. Here, we predict that the PSH can be intrinsically achieved on a two-dimensional (2D) group-IV monochalcogenide $MX$ monolayer, a new class of the noncentrosymmetric 2D materials having in-plane ferroelctricity. Due to the $C_{2v}$ point group symmetry in the $MX$ monolayer, a unidirectional spin configuration is preserved in the out-of-plane direction and thus maintains the PSH that is similar to the [110] Dresselhaus model in the [110]-oriented quantum well. Our first-principle calculations on various $MX$ ($M$: Sn, Ge; $X$: S, Se, Te) monolayers confirmed that such typical spin configuration is observed, in particular, at near the valence band maximum where a sizable spin splitting and a substantially small wavelength of the spin polarization are achieved. Importantly, we observe reversible out-of-plane spin orientation under opposite in-plane ferroelectric polarization, indicating that an electrically controllable PSH for spintronic applications is plausible.     
\end{abstract}

\pacs{Valid PACS appear here}
\maketitle

\section{INTRODUCTION}
Recent development of spintronics relies on the new pathway for exploiting electron's spin in semiconductors by utilizing the effect of spin-orbit coupling (SOC)\cite{Nitta,Manchon}. In a system with lack of inversion symmetry, the SOC induces an effective magnetic field or known as a spin-orbit field (SOF) acting on spin, so that the effective SOC Hamiltonian can be expressed as 
\begin{equation}
\label{1}
H_{\text{SOC}}=\vec{\Omega}(\vec{k})\cdot \vec{\sigma}=\alpha (\hat{E}\times \vec{k})\cdot \vec{\sigma},
\end{equation}
where $\vec{\Omega}$ is the SOF vector, $\vec{k}$ is the wave vector representing momentum of electrons, $\vec{\sigma}=(\sigma_{x}, \sigma_{y}, \sigma_{z})$ is the Pauli matrices vector, and $\alpha$ is the strength of the SOC that is proportional to magnitude of local electric field $\vec{E}$ induced by the crystal inversion asymmetry. Since the SOF is odd in the electron's wave vector $\vec{k}$, as was firstly demonstrated by Dresselhauss \cite {Dresselhauss} and Rashba \cite {Rashba}, the SOC lifts Kramers' spin degeneracy and leads to a complex $\vec{k}$-dependent spin configuration of the electronic bands. In particular interest is driven due to a possibility to manipulate this spin configuration by using an external electric field to create non-equilibrium spin polarization \cite{Kuhlen}, leading to various important phenomena such as spin Hall effect \cite{Qi}, spin galvanic effect \cite{Ganichev}, and spin ballistic transport \cite{Lu}, thus offering for realization of spintronics device such as spin-field effect transistor (SFET)\cite {Datta}. 

From practical perspective, materials having strong Rashba SOC have generated significant interest since they allow for electrostatic manipulation of the spin states \cite{Nitta,Chuang}, paving the way towards non-charge-based computing and information processing \cite{Manchon}. However, the strong SOC is also known to induce the undesired effect of causing spin decoherence \cite{Dyakonov}, which plays an adverse role in the spin lifetime. In a diffusive transport regime, impurities and defects act as scatters which change the momentum of electron and simultaneously randomize the spin due to momentum-dependent SOF, leading to the fast spin decoherence through the Dyakonov-Perel (DP) mechanism of spin-relaxation\cite {Dyakonov}.This process induces spin dephasing and a loss of the spin signal, such that the spin lifetime significantly reduces, thus limiting the performance of potential spintronic devices.

A possible way to overcome this obstacle is to eliminate the problem of the spin dephasing by suppressing the DP spin relaxation. This can be achieved, in particular, by designing a structure where the SOF orientation is enforced to be unidirectional, preserving a unidirectional spin configuration in the momentum space. In such situation, electron motion together with the spin precession around the unidirectional SOF leads to a spatially periodic mode of the spin polarization known as persistent spin helix (PSH)\cite{Bernevig,Schliemann}. The corresponding spin wave mode protects the spins of electron from the dephasing due to $SU(2)$ spin rotation symmetry, which is robust against spin-independent scattering and renders an extremely long spin lifetime \cite{Bernevig,Altmann}. Previously, the PSH has been demonstrated on various [001]-oriented semiconductors quantum well (QW) \cite{Koralek,Walser,Schonhuber,Ishihara,Kohda,Sasaki} having equal strength of the Rashba and Dresselhauss SOC, or on [110]-oriented semiconductor QW \cite{Chen} in which the SOC is described by the [110] Dreseelhauss model. Here, for the former, the spin configurations are enforced to be unidirectional in the in-plane [110] direction, whereas for the latter, they are oriented in the out-of-plane [001] direction. Similar to the [110]-oriented QW, the PSH state has recently been reported for LaAlO$_{3}$/SrTiO$_{3}$ interface \cite{Yamaguchi}, ZnO [10-10] surface \cite{Absor1}, halogen-doped SnSe monolayer \cite{Absor2}, and WO$_{2}$Cl$_{2}$ monolayer\cite{Ai}. Although the PSH has been widely studied on various QW systems \cite {Koralek,Walser,Schonhuber,Ishihara,Kohda,Sasaki}, it is practically non-trivial due to the stringent conditions for fine-tuning the Rashba and Dresselhaus SOCs. Therefore, it would be desirable to find a new class of material which intrinsically supports the PSH.

In this paper, we show that the PSH can be intrinsically achieved on a two-dimensional (2D) group-IV monochalcogenide $MX$ monolayer, a new class of noncentrosymmetric 2D materials having in-plane ferroelctricity\cite{Fei,Lopez,Kaloni,WWan,Hanakata}. On the basis of density-functional theory (DFT) calculations on various $MX$ ($M$: Sn, Ge; $X$: S, Se, Te) monolayers, supplemented with symmetry analysis, we find that a unidirectional spin orientation is preserved in the out-of-plane direction, yielding a PSH that is similar to the [110] Dresselhaus model in the [110]-oriented QW. Such typical spin configuration is observed, in particular, at near the valence band maximum, having a sizable spin splitting and small wavelength of the spin polarization. More interestingly, we observe reversible out-of-plane spin orientation under opposite in-plane ferroelectric polarization, suggesting that an electrically controllable PSH is achievable, which is useful for spintronic applications.

\section{Computational details}

We performed first-principles calculations by using DFT within the generalized gradient approximation (GGA) \cite {Perdew} implemented in the OpenMX code \cite{Openmx}. Here, we adopted norm-conserving pseudopotentials \cite {Troullier} with an energy cutoff of 350 Ry for charge density. The $12\times12\times1$ k-point mesh was used. The wave functions were expanded by linear combination of multiple pseudoatomic orbitals generated using a confinement scheme \cite{Ozaki,Ozakikino}, where two $s$-, two $p$-, two $d$-character numerical pseudo-atomic orbitals were used. The SOC was included in the DFT calculations by using $j$-dependent pseudopotentials \citep{Theurich}. The spin textures in the momentum space were calculated using the spin density matrix of the spinor wave functions obtained from the DFT calculations as we applied recently on various 2D materials \citep{Absor1,Absor2,Absor3,Absor4,Absor5,Absor6}.

\begin{table}[ht!]
\caption{Structural-related parameters corresponding to the band gap of the $MX$ monolayer. $a$ and $b$ (in \AA) represent the lattice parameters in the $x$- and $y$-directions, respectively. $d_{1}$ and $d_{2}$ (in \AA) indicate the bondlength between the $M$ ($M$: Sn, Ge) and $X$ ($X$: S, Se, Te) atoms in the in-plane and out-of-plane directions, respectively. $E_{g}$ (in eV) represents the energy gap where the star (*) indicates direct band gap.} 
\centering 
\begin{tabular}{c c c c c c} 
\hline\hline 
  $MX$ monolayer & $a$ (\AA)  & $b$ (\AA)  & $d_{1}$ (\AA) & $d_{2}$ (\AA) & $E_{g}$ (eV) \\ 
\hline 
SnS & 4.01  & 4.39  & 2.71 & 2.63  & 1.38 \\
SnSe &  4.34  & 4.49   & 2.89 & 2.7   & 0.98*  \\
SnTe &   4.54  & 4.58   & 3.31 & 3.04  &  0.85\\
GeS &   3.68  & 4.40   & 2.53 & 2.46  & 1.45 \\
GeSe &   3.99  & 4.26   & 2.72 &  2.57 & 1.10* \\
GeTe &   4.27  & 4.47  & 2.95 &  2.81 & 0.92 \\
\hline\hline 
\end{tabular}
\label{table:Table 1} 
\end{table}

\begin{figure}
	\centering		
	\includegraphics[width=0.65\textwidth]{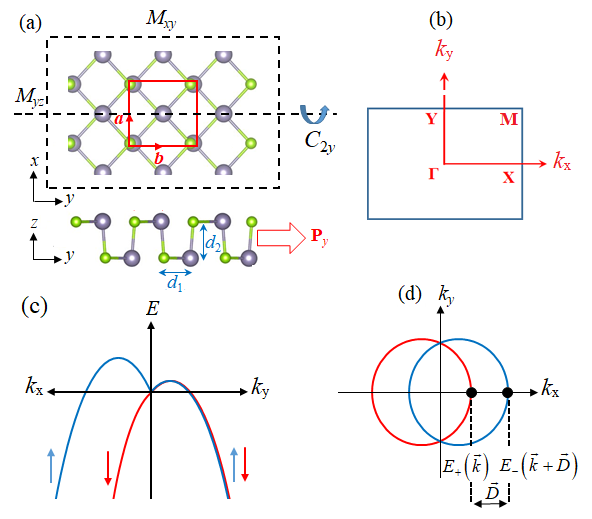}
	\caption{ (a) Atomic structure of the $MX$ monolayer corresponding to it symmetry operations. Black and green balls represent the $M$ ($M$: Sn, Ge) and $X$ ($X$: S, Se, Te) atoms, respectively. The unit cell of the crystal is indicated by red lines characterized by $a$ and $b$ lattice parameters in the $x$ and $y$ directions. $d_{1}$ and $d_{2}$ represent bondlength between the $M$ ($M$: Sn, Ge) and $X$ ($X$: S, Se, Te) atoms in the in-plane and out-of-plane directions, respectively. (b) First Brillouin zone of the $MX$ monolayer characterized by high symmetry $\vec{k}$ points ($\Gamma$, Y, M, X) are shown. (c) Spin-split bands induced by the SOC and $C_{2v}$ point group symmetry and (d) the corresponding Fermi contours in the momentum space are schematically shown. Here, the Fermi contours are characterized by two Fermi loops shifted by the wave vector $\vec{D}$, exhibiting a unidirectional spin configuration in the out-of-plane direction. The red and blue lines (or arrows) represent positive and negative spins, respectively, in the out-of-plane directions.}
	\label{figure:Figure1}
\end{figure}

In our DFT calculations, we considered ferroelectric phase of the $MX$ monolayer having black phosporene-type structure \cite{Gomes,Appelbaum}. The minimum energy pathways of ferroelectric transitions were calculated using nudged elestic band (NEB) method\cite{NEB} based on the interatomic forces and total energy obtained from DFT caclulations. The Ferroelectric polarization was calculated using Berry phase approach \cite{BP}, where both electronic and ionic contributions were considered. We used a periodic slab to model the $MX$ monolayer, where a sufficiently large vacuum layer (20 \AA) is applied in order to avoid interaction between adjacent layers. We used the axes system where layers are chosen to sit on the $x-y$ plane, while the $x$ axis is taken to be parallel to the puckering direction [Fig. 1(a)]. The geometries were fully relaxed until the force acting on each atom was less than 1 meV/\AA. The optimized structural-related parameters are summarized in Table 1, where overall are in good agreement with previously reported data \cite{LXu,Gomes,WWan}.

\section{Results and Discussion}

\subsection{Symmetry-protected PSH state in $MX$ monolayer}

To predict the PSH state in the $MX$ monolayer, we firstly derive an effective low energy Hamiltonian by using symmetry analysis. As shown in Fig. 1(a), the crystal structures of the $MX$ ML has black phosporene-type structures where the symmetry group is isomorphic to $C_{2v}^{7}$ or $Pmn2_{1}$ space group \cite{Appelbaum,Gomes}. There are four symmetry operations in the crystal lattice of the $MX$ monolayer [Fig. 1(a)]: (i) identity operation $E$; (ii) twofold screw rotation $\bar{C}_{2y}$ (twofold rotation around the $y$ axis, $C_{2y}$, followed by translation of $\tau=a/2,b/2$), where $a$ and $b$ is the lattice parameters along $\vec{a}$ and $\vec{b}$ directions, respectively; (iii) glide reflection $\bar{M}_{xy}$ (reflection with respect to the $xy$ plane followed by translation $\tau$); and (iv) reflection $M_{yz}$ with respect to the $yz$ plane. The effective $\vec{k}\cdot \vec{p}$ Hamiltonian can be constructed by taking into account all symmetry operations in the little group of the wave vector in the reciprocal space.

\begin{table}[ht!]
\caption{Transformation rules for the in-plane wave vector components ($k_{x}$, $k_{y}$) and spin Pauli matrices $(\sigma_{x}, \sigma_{y}, \sigma_{z})$ under the considered point-group symmetry operations. Time-reversal symmetry, implying a reversal of both spin and momentum, is defined as $T=i\sigma_{y}K$, where $K$ is the complex conjugation, while the point-group operations
are defined as $\hat{C}_{2y}=i\sigma_{y}$, $\hat{M}_{yz}=i\sigma_{x}$, and $\hat{M}_{xy}=i\sigma_{z}$.} 
\centering 
\begin{tabular}{c c c} 
\hline\hline 
  Symmetry operation & $(k_{x}, k_{y})$   & $(\sigma_{x}, \sigma_{y}, \sigma_{z})$ \\ 
\hline 
$\hat{T}=i\sigma_{y}K$   &  $(-k_{x}, -k_{y})$  &  $(-\sigma_{x}, -\sigma_{y}, -\sigma_{z})$    \\  
$\hat{C}_{2y}=i\sigma_{y}$   &  $(-k_{x}, k_{y})$  &  $(-\sigma_{x}, \sigma_{y}, -\sigma_{z})$    \\ 
$\hat{M}_{yz}=i\sigma_{x}$   &  $(-k_{x}, k_{y})$  &  $(\sigma_{x}, -\sigma_{y}, -\sigma_{z})$    \\
$\hat{M}_{xy}=i\sigma_{z}$   &  $(k_{x}, k_{y})$  &  $(-\sigma_{x}, -\sigma_{y}, \sigma_{z})$   \\
\hline\hline 
\end{tabular}
\label{table:Table 2} 
\end{table}

Let $Q$ be a high symmetry point in the first Brillouin zone (FBZ) where a pair of spin-degenerate eigen states exsist in the valence band maximum (VBM) or conduction band minimum (CBM). This degeneracy appers due to time reversal symmetry $T$ for which the condition that $\vec{Q}=-\vec{Q}+\vec{G}$ is satisfied, where $\vec{G}$ is the 2D reciprocal-lattice vector. Such points are located at the center of the FBZ ($\Gamma$ point), or some points that are located at the boundary of the FBZ such as $X$, $Y$, and $M$ points for a primitive rectangular lattice [Fig. 1(b)]. The band dispersion around the $Q$ point can be deduced by identifying all symmetry-allowed terms so that $O^{\dagger}H(k)O=H(k)$ is obtained, where $O$ denotes all symmetry operations belonging to the little group of the $Q$ point, supplemented by time-reversal symmetry $T$.

For simplicity, let we assume that the little group of the wave vector $\vec{k}$ at the $Q$ point belongs to the $C_{2v}$ point group similar to that of the crystal in the real space. Therefore, the wave vector $\vec{k}$ and spin vector $\vec{\sigma}$ can be transformed according to the symmetry operation $O$ in the $C_{2v}$ point group and time reversal symmetry $T$. The corresponding transformation for the $\vec{k}$ and $\vec{\sigma}$ are listed in Table II. Collecting all terms which are invariant with respect to the symmetry operation, we obtain the following effective Hamiltonian up to third order correction of $k$\cite{Schliemann}:
\begin{equation}
\label{2}
\begin{split}
H & = E_{0}(k)+\alpha k_{x}\sigma_{z}+ (\alpha^{'}k_{y}^{2} k_{x}+\alpha^{"}k_{x}^{3})\sigma_{z}\\
& = E_{0}(k)+\alpha^{(1)}k\cos\theta \sigma_{z}+\alpha^{(3)}k\cos(3\theta) \sigma_{z},
\end{split}
\end{equation}
where $E_{0}(k)=\hbar^{2}(k_{x}^{2}+k_{y}^{2})/2m^{*}$ is the nearly free electron/hole energy, $\alpha^{(1)}$ defined as $\alpha^{(1)}=\alpha+(k^{2}/4)(\alpha^{'}+3\alpha^{"})$ is originated from the contribution of the $k$-linear parameter $\alpha$ and the correction provided by the third order parameters ($\alpha^{'}$ and $\alpha^{"}$), $\alpha^{(3)}$ corresponds to the third order parameters by the relation $\alpha^{(3)}=(1/4)[\alpha^{'}-\alpha^{"}]k^{2}$, and $\theta$ is the angle of the momentum $\vec{k}$ with respect to the $x$-axis defined as $\theta=\cos^{-1}(k_{x}/k)$. Solving the eigenvalue problem involving the Hamiltonian of Eq. (\ref{2}) yields split-split energy dispersions: 
\begin{equation}
\label{3}
E_{\pm}=E_{0}(k) \pm [\alpha^{(1)}\cos\theta+\alpha^{(3)}\cos(3\theta)]k.
\end{equation}
These dispersions are schematically illustrated in Fig. 1(c) showing a highly anisotropic spin splitting.

Since the Hamiltonian of Eq. (\ref{2}) is only coupled with $\sigma_{z}$, neglecting all the cubic terms leads to the $SU(2)$ symmetry of the Hamiltonian \cite{Schliemann,Bernevig}, 
\begin{equation}
\label{4}
H=E_{0}(k)+\alpha k_{x}\sigma_{z},
\end{equation}
with the energy dispersions, 
\begin{equation}
\label{5}
E_{\pm}=E_{0}(k) \pm \alpha k_{x}.
\end{equation}
Importantly, these dispersions have the shifting property: $E_{+}(\vec{k})=E_{-}(\vec{k}+\vec{D})$, where $\vec{D}=2m^{*}\alpha(1,0,0)/\hbar^{2}$ is the shifting wave vector. As a result, constant-energy cut shows two Fermi loops whose centers are displaced from their original point by $\mp \vec{D}$ as schematically shown in Fig. 1(d).

Since the $z$ component of the spin operator $S_{z}$ commutes with this Hamiltonian of Eq. (\ref{4}), $[S_{z},H]=0$, the spin operator $S_{z}$ is a conserved quantity. Here, expectation value of the spin $\left\langle S\right\rangle$ only has the out-of-plane component: $(\left\langle S_{x}\right\rangle,\left\langle S_{y}\right\rangle,\left\langle S_{z}\right\rangle)_{\pm} =\pm (\hbar/2)(0,0,1)$ at any wave vector $\vec{k}$ except for $k_{x}=0$, resulting in the unidirectional out-of-plane spin configuration in the momentum space [Fig. 1(d)]. In such situation, the unidirectional out-of-plane SOF is achieved, implying that the electron motion accompanied by the spin precession around the SOF form a spatially periodic mode of the spin polarization, yielding the PSH that is similar to the [110] Dresselhaus model \cite{Bernevig} as recently demonstrated on the [110]-oriented semiconductor QW\citep{Chen}. 

In the next section, we discuss our results from the first-principles DFT calculations on various $MX$ ($M$: Sn, Ge; $X$: S, Se, Te) monolayers to confirm the above predicted PSH.

\subsection{DFT analysis of $MX$ monolayer}

Figure 2 shows the electronic band structures of various $MX$ ($M$: Sn, Ge; $X$: S, Se, Te) monolayers  calculated along the selected $\vec{k}$ paths in the FBZ corresponding to the density of states (DOS) projected to the atomic orbitals. Without including the SOC, it is evident that there are two equivalent extrema valleys characterizing the VBM and CBM located at the points that are not time reversal invariant. Consistent with previous calculations\cite{WWan,Gomes,LXu}, the $MX$ monolayers show indirect band gap (except for $M$Se monolayer), where the VBM and CBM are located along the $\Gamma$-$Y$ and $\Gamma$-$X$ lines, respectively. Overall, the calculated band gap [see Table I] is in a good agreement with previous results under GGA-PBE level \cite{Gomes,LXu}. Our calculated results of the DOS projected to the atomic orbitals confirmed that the $M$-$s$ and $X$-$p$ orbitals contributes dominantly to the VBM, while the CBM is mainly originated from the contribution of the $M$-$p$ and $X$-$s$ orbitals. 

\begin{figure*}
	\centering		
	\includegraphics[width=1.0\textwidth]{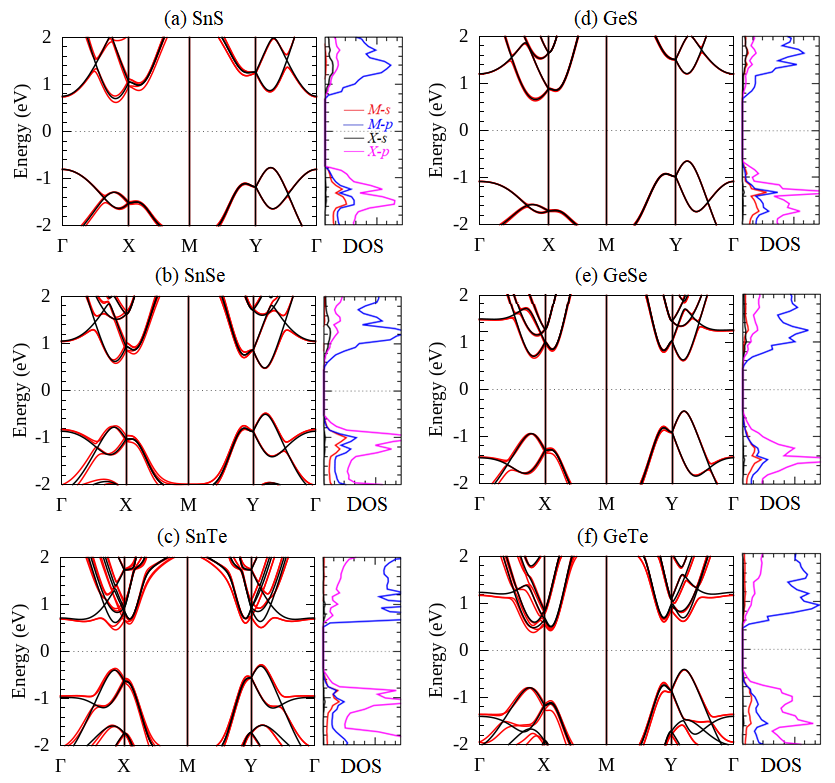}
	\caption{ (a) Electronic band structures of the $MX$ monolayers corresponding to density of state projected to the atomic orbitals for: (a) SnS, (b) SnSe, (c) SnTe, (d) GeS, (e) GeSe, and (f) GeTe. The balck and red lines show the calculated band structures without and with the SOC, respectively.}
	\label{figure:Figure2}
\end{figure*}

Turning the SOC strongly modifies the electronic band structures of the $MX$ monolayers [Fig. 2]. Importantly, a sizable splitting of the bands produced by the SOC is observed at some high symmetry $\vec{k}$ points and along certain $\vec{k}$ paths in the FBZ. This splitting is especially pronounced around the X and Y points near both the VBM and CBM. However, there are special high-symmetry lines and points in the FBZ where the splitting is zero. This is in particular, the case for $\Gamma$-$Y$ line, where the wave vector $\vec{k}=(0,k_{y},0)$ is parallel to the ferroelectric polarization along the $y$ direction. 

To analyze the properties of the spin splitting, we consider SnTe monolayer as a representative example of the $MX$ monolayer. Here, we focus our attention on the bands near the VBM (including spin) around the Y point due to the large spin splitting as highligted by the blue lines in Fig. 3(a). Without the SOC, it is clearly seen from the band dispersion that fourfold degenerate state is visible at the $Y$ point [Fig. 3(b)]. Taking into account the SOC, this degeneracy splits into two pair doublets with the splitting energy of $\Delta E_{Y}=9.2$ meV [Fig. 3(c)]. Although these doublets remain at the $\vec{k}$ along the $\Gamma$-$Y$ line, they split into a singlet when moving away along the $Y$-$M$ line, yielding a highly anisotropic spin splitting. 

\begin{figure*}
	\centering		
	\includegraphics[width=1.0\textwidth]{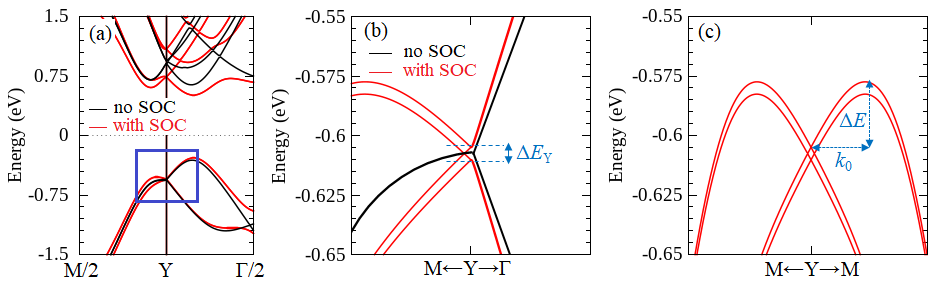}
	\caption{ (a) Energy band dispersion of SnTe monolayer along the $M$-$Y$-$\Gamma$ lines calculated without (black lines) and with (red lines) the SOC are shown. (b) Zoom-in the energy dispersion near the VBM closed to the $Y$ point along $M$-$Y$ and $Y$-$\Gamma$ lines as highlighted by the blue lines in Fig. 3(a). (c) Spin splitting properties of the bands around the $Y$ point along the $M$-$Y$-$M$ lines  characterized by: (i) splitting energy ($\Delta E$), i.e., different energy between the VBM along $Y$-$M$ line and the energy band at the $Y$ point, and (ii) momentum offset ($k_{0}$). }
	\label{figure:Figure3}
\end{figure*}

To clarify the origin of the anisotropic splitting around the $Y$ point near the VBM, we discuss our system based on the symmetry argument. At the $Y$ point, the little group of the wave vector $\vec{k}$ belongs to the $C_{2v}$ point group\cite{Appelbaum}. As previously mentioned that the $C_{2v}$ point group contains the $C_{2y}$ rotation symmetry around the $y$-axis. Applying the $C_{2y}$ rotation twice to the Bloch wave function, we have $C_{2y}^{2}\psi_{k}=e^{ik_{y}b} \psi_{k}$, thus we obtain that $C_{2y}^{2}=e^{ik_{y}b}$. We further define an antiunitary symmetry operator, $\Theta=C_{2y}T$, so that $\Theta^{2}=C_{2y}^{2}T^{2}=-e^{ik_{y}b}$ for spin half system. Therefore, at the $Y$ point ($k_{y}=\pi/b$), we find that $\Theta^{2}=-1$, thus the Bloch states $(\psi_{k},\Theta\psi_{k})$ are double degenerate. 

\begin{figure*}
	\centering		
	\includegraphics[width=0.5\textwidth]{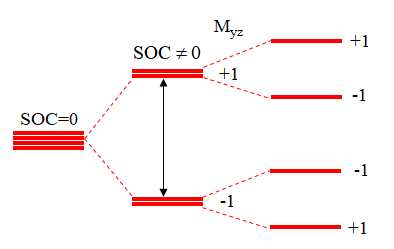}
	\caption{Schematic view of the energy level around the $Y$ point near the VBM. The SOC splits the states into two doublets with eigenvalues of $M_{yz}=\pm 1$, which are further splitted into a singlet with sign-reversed expectetion values of spin.}
	\label{figure:Figure4}
\end{figure*}

In addition, there is also $M_{yz}$ mirror symmetry in the $C_{2v}$ point group, which commutes with Hamiltonian of the crystal, $[M_{yx},H]=0$. By operating $M_{yz}$ symmetry to the Bloch states, we find that $M_{yz}^{2}=-e^{-ik_{y}b}$. Accordingly, the Bloch states can be labelled using the $M_{yz}$ eigenvalues, i.e., $M_{yz}\ket{\psi_{k}^{\pm}}=\pm ie^{ik_{y}b/2}\ket{\psi_{k}^{\pm}}$. Here, for the $Y$ point ($k_{y}=\pi/b$), we find that $M_{yz}^{2}=1$, thus we obtain $M_{yz}\psi_{Y}^{\pm}=\pm \psi_{Y}^{\pm}$ and $M_{yz}\Theta\psi_{Y}^{\pm}=\pm \Theta\psi_{Y}^{\pm}$. Therefore, there are two conjugated doublets at the $Y$ point, $(\psi_{Y}^{+},\Theta\psi_{Y}^{+})$ or $(\psi_{Y}^{-},\Theta\psi_{Y}^{-})$, which is distinguished by the $M_{yz}$ eigenvalues as schematically shown in Fig. 4. These conjugated doublets are preserved along the $\Gamma$-$Y$ line but they split into singlet when moving to the $Y$-$M$ line, which are protected by the $M_{yz}$ and $C_{2y}$ symmtery operations. As a result, the strong anisotropic splitting is achieved, which is in fact consistent well with our DFT results shown in Fig. 3(c).

\begin{figure}
	\centering		
	\includegraphics[width=1.0\textwidth]{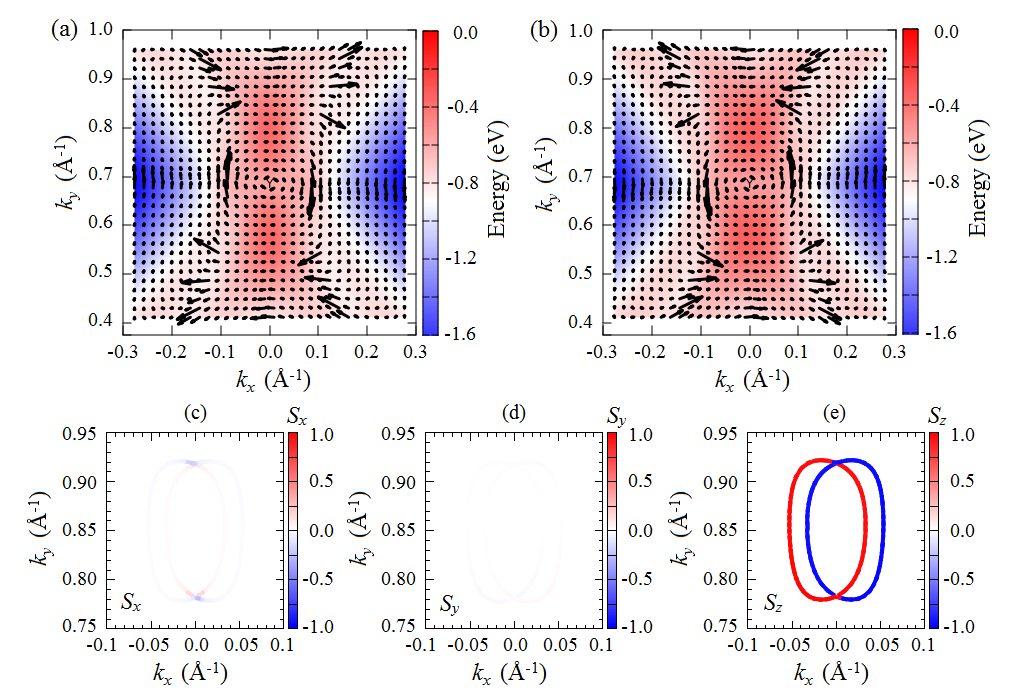}
	\caption{ Energy profiles of the spin textures calculated around the $Y$ point near the VBM for: (a) upper and (b) lower bands. The colours scale in Fig. 5(a)-(b) indicate the energy band near the VBM. Constant energy contours corresponding to a cut at 1 meV below the VBM characterized by (c)   $S_{x}$, (d) $S_{y}$, and (e) $S_{z}$ components of the spin distribution are shown. The colours scale in Fig. 5(c)-(e) show the modulus of the spin polarization.}
	\label{figure:Figure5}
\end{figure}

To further demonstrate the nature of the observed anisotropic splitting around the $Y$ point near the VBM, we show in Figs. 5(a) and 5(b) the energy profiles of the spin textures for the upper and lower bands, respectively. It is found that a complex pattern of the spin polarization is observed around the $Y$ point, which is remarkably different either from Rashba- and Dresselhaus-like spin textures. This is in contrast to the widely studied 2D materials such as PtSe$_{2}$ \cite{Absor3,WYao}, BiSb \cite{QLiu}, LaOBiS$_{2}$ \cite{SSingh}, and polar transition metal dichalcogenide \cite{Absor5,Absor6}, where the Rashba-like spin textures are identified. In particular, we observe a unifrom spin polarization close to the VBM, which persists in a region located at about 0.1 \AA$^{-1}$ from $Y$ point along the $Y$-$M$ and $Y$-$\Gamma$ lines [see the region with red colour in Fig. 5(a)-(b)]. By carefully analyzing the spin textures measured at the constant energy cut of 1 meV below the VBM, we confirmed that this peculiar spin polarization is mostly dominated by the out-of-plane component $S_{z}$ [Fig. 5(e)] rather than the in-plane ones ($S_{x}$, $S_{y}$) [Fig. 5(c)-(d)], leading to the unidirectional out-of-plane spin textures. On the other hand, the constant-energy cut also induces the Fermi lines characterized  by the shifted two circular loops along the $Y$-$M$ ($k_{x}$) direction and the degenerated nodal point along the $Y$-$\Gamma$ ($k_{y}$) direction. Both the spin textures and Fermi lines are agree well with our $\vec{k}\cdot\vec{p}$ Hamiltonian model derived from the symmetry analysis. Since the spin textures are uniformly oriented in the out-of-plane direction, the unidirectional out-of-plane SOF is achieved, maintaining the PSH that is similar to the [110] Dresselhauss model\citep{Bernevig}. Therefore, it is expected that the DP mechanism of the spin relaxation is suppressed, potentially ensurring to induce an extremely long spin lifetime.
 
\begin{table}[ht!]
\caption{Spin splitting parameter  $\alpha$ (in eV\AA) and the wavelength of the spin polarization $\lambda$ (in nm) for the  selected PSH materials.} 
\centering 
\begin{tabular}{c c c c} 
\hline\hline 
  Systems & $\alpha$ (eV\AA)  & $\lambda$ (nm)  & Reference \\ 
\hline 
$MX$ monolayer    &    &  &   \\ 
SnS    & 0.09   & 1.5$\times10^{2}$ & This work  \\ 
SnSe   & 0.74  & 44.85 & This work  \\ 
SnTe    & 1.20   & 7.13 & This work  \\ 
GeS    & 0.071   & 8.9$\times10^{2}$ & This work  \\ 
GeSe    & 0.57   & 91.84 & This work  \\ 
GeTe    & 1.67   & 1.82 & This work  \\ 	
Interface    &    &  &   \\ 
GaAs/AlGaAs & (3.5-4.9)$\times 10^{-3}$  & (7.3-10) $\times10^{3}$& Ref.\cite{Walser} \\
               & 2.77 $\times10^{-3}$ & 5.5$\times10^{3}$ & Ref.\cite{Schonhuber} \\
InAlAs/InGaAs & 1.0 $\times10^{-3}$&      & Ref.\cite{Ishihara}\\
              & 2.0 $\times10^{-3}$&      & Ref.\cite{Sasaki}\\
LaAlO$_{3}$/SrTiO$_{3}$ & 7.49 $\times10^{-3}$ & 0.098$\times10^{2}$  & Ref.\cite{Yamaguchi}\\
Surface    &    &  &   \\               
ZnO(10-10) surface & 34.78 $\times10^{-3}$& 1.9$\times10^{2}$ & Ref.\cite{Absor1}\\
Bulk    &    &  &   \\ 
BiInO$_{3}$ &1.91 & 2.0 & Ref.\cite{LLTao}\\
2D monolayer    &    &  &   \\ 
Halogen-doped SnSe & 1.6-1.76 & 1.2-1.41 & Ref.\cite{Absor2}\\
WO$_{2}$Cl$_{2}$ &  0.9     &      & Ref.\cite{Ai}\\
\hline\hline 
\end{tabular}
\label{table:Table 3} 
\end{table}

For a quantitative analysis of the above mentioned spin splitting, we here calculate the strength of the spin splitting by evaluating the band dispersions along the $Y$-$M$ and the $Y$-$\Gamma$ directions near the VBM in term of the effective $\vec{k}\cdot\vec{p}$ Hamiltonian model given in Eq. (\ref{2}). Here, according to Eq. (\ref{3}), the spin-splitting energy ($E_{\text{Split}}=E_{+}-E_{-})$) can be formulated as 
\begin{equation}
\label{6}
E_{\text{Split}}=2k[(\alpha+(k^{2}/4)(\alpha^{'}+3\alpha^{"}))\cos\theta+(k^{2}/4)(\alpha^{'}-\alpha^{"})\cos(3\theta)].
\end{equation}
The parametrs $\alpha$, $\alpha^{'}$, and $\alpha^{"}$ can be calculated by numerically fitting of Eq. (\ref{6}) to the spin splitting energy along the $Y$-$M$ ($k_{x}$) and the $Y$-$\Gamma$ ($k_{y}$) directions obtained from our DFT results, and find that $\alpha=1.23$ eV\AA, $\alpha^{'}=0.0014$ eV\AA$^{3}$, and $\alpha^{"}=0.0027$ eV\AA$^{3}$. It is clearly seen that the obtained value of the cubic term parameters ($\alpha^{'}$, $\alpha^{"}$) is too small compared with that of the linear term parameter $\alpha$, indicating that the contribution of the higher order correction is not essential. On the other hand, by using the energy dispersion of Eq. (\ref{5}), we also obtain the linear term parameter $\alpha$ through the relation $\alpha=2E_{R}/k_{0}$, where $E_{R}$ and $k_{0}$ are the shifting energy and the wave vector as illustrated in Fig. 3(c). This revealed that the calculated value of $\alpha$ is 1.20 eV\AA, which is fairly agree with that obtained from the higher order correction model. Since the spin-splitting is dominated by the linear term, ignoring the higher order correction preserves the $SU(2)$ symmetry of the Hamiltonian, thus maintaining the PSH as we expected. 

It is important to noted here that the PSH predicted in the present system should ensure that a spatially periodic mode of spin polarization is achieved. The corresponding spin wave mode is characterized by the wavelength of the spin polarization defined as \cite{Bernevig} $\lambda=(\pi\hbar^{2})/(m^{*}\alpha)$, where $m^{*}$ is the hole effective mass. Here, the effective mass $m^{*}$ can be evaluated by fitting the sum of the band dispersions along the $Y$-$M$ direction in the VBM. Here, we find that $m^{*}=0.056m_{0}$, where $m_{0}$ is the free electron mass, which is in a good agreement with previous result reported by Xu et. al.\cite{LXu} The resulting wavelength $\lambda$ is 7.13 nm, which is typically on the scale of the lithographic dimension used in the recent semiconductor industry \cite{Fiori}. 

We summarize the calculated results of the $\alpha$ and $\lambda$ in Table III and compare the results with a few selected PSH materials from previously reported data. It is found that the calculated value of $\alpha$ in various $MX$ monolayer is much larger than that observed on various QWs such as GaAs/AlGaAs \cite{Walser,Schonhuber} and InAlAs/InGaAs \cite{Sasaki,Ishihara}, ZnO (10-10) surface\cite{Absor1}, and strained LaAlO3/SrTiO3 (001) interface \cite{Yamaguchi}. However, this value is comparable with those observed on the bulk BiInO$_{3}$ \cite{LLTao}, halogen-doped SnSe monolayer\cite{Absor2}, and WO$_{2}$Cl$_{2}$ monolayer\cite{Ai}. The associated spin-splitting parameters are sufficient to support room temperature spintronics functionality. On the other hand, we observed small wavelength $\lambda$ (in nm scale) of the spin polarization, which is in fact two order less than that observed on the GaAs/AlGaAs QW \cite{Walser,Schonhuber}, rendering that the present system is promising for nanoscale spintronics devices.

\begin{figure}
	\centering		
	\includegraphics[width=0.7\textwidth]{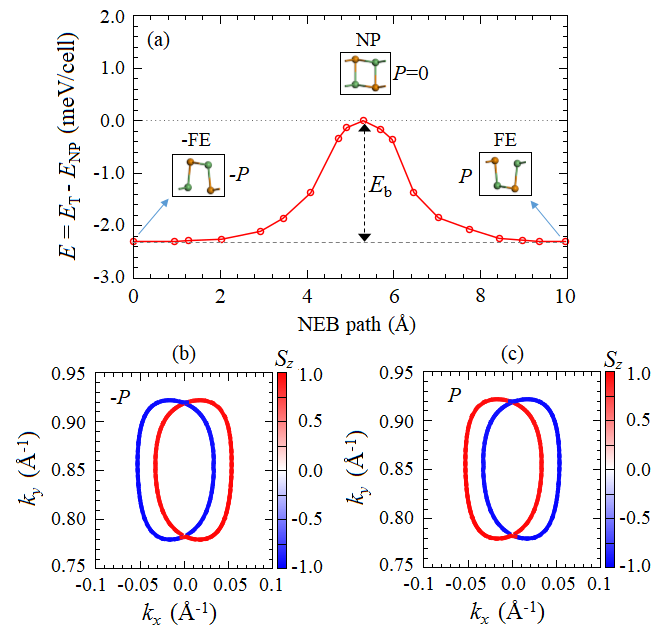}
	\caption{(a) Nudeged elastic band calculation for the polarization switching process through centrosymmetric (paraelectric) structures in SnTe monolayer. Two ferroelectric structures (FE) in the ground state with opposite direction of the electric polarization and a paraelectric structure with zero electric polarization (NP) are shown. $E_{b}$ is the barrier energy defined as the energy different between the total energy of the ferroelectric and paraelectric structure. Reversible out-of-plane spin orientation in SnTe monolayer calculated at 1 meV below the VBM for the ferroelectric structure with opposite polarization: (b) -P  and (c) P. }
	\label{figure:Figure6}
\end{figure}

Now, we discuss our prediction of the PSH in correlated to the ferroelectricity in the $MX$ monolayer. As previously mentioned that the $MX$ monolayer posses in-plane ferroelectricity \cite{Fei,Lopez,Kaloni,WWan,Hanakata}, which is induced by the in-plane atomic distortion in the real space of the crystal [see Fig. 1(a)]. Therefore, a substantial electric polarization in the in-plane direction is established. For instant, our Berry phase calculation \cite{BP} on SnTe monolayer revealed that the magnitude of the in-plane electric polarization is 13.8 $\mu \text{C}/\text{cm}^{2}$ when an effective thickness of 1 nm for monolayer is used, which is in a good agreement with previous result\cite{WWan}. Importantly, we predict the feasibility of polarization switching in SnTe monolayer by analyzing the minimum energy pathway of ferroelectric transition calculated using NEB method \cite{NEB}. As shown in Fig. 6(a), we find that the calculated barrier energy for polarization switching process is 2.26 meV/cell in SnTe monolayer. This value is comparable to those of the 2D ferroelectric reported in previous work\cite{Fei,Ai}, but is much smaller than that in conventional ferroelectric BaTiO$_{3}$\cite{Ba}, suggesting that a switchable in-plane ferroelectric polarization is plausible. In deed, polarization switching in various $MX$ monolayers by using an external electric field or strain effects has recently been  reported \cite{Hanakata}.

By switching the in-plane ferroelectric polarization $\vec{P}$ in $MX$ monolayer, e.g., by applying an external electric field, a fully reversal of the out-of-plane spin orientation can be expected. This is due to the fact that switching the in-plane ferroelectric polarization from $\vec{P}$ to $-\vec{P}$ is equivalent to the space inversion operation which changes the wave vector from $\vec{k}$ to $-\vec{k}$, but preserves the spin vector $\vec{\sigma}$ \cite{DSante,MKim}. Suppose that $\ket{\psi_{\vec{P}}(\vec{k})}$ is the Bloch state of the crystal with ferroelectric polarization $\vec{P}$. Under the space inversion operation $I$, both the polarization $\vec{P}$ and the wave vector $\vec{k}$ are reversed so that $I\ket{\psi_{\vec{P}}(\vec{k})}=\ket{\psi_{-\vec{P}}(\vec{-k})}$. However, application of the time reversal symmetry $T$ reverses only the $\vec{k}$, while the $\vec{P}$ remains unchanged, leading to the fact that $TI\ket{\psi_{\vec{P}}(\vec{k})}=\ket{\psi_{-\vec{P}}(\vec{k})}$. The expectation values of spin operator $\left\langle S\right\rangle$ can now be calculated by
\begin{equation}
\label{7}
\begin{split}
\expval{S}_{-\vec{P},\vec{k}} & =\bra{\psi_{-\vec{P}}(\vec{k})}S\ket{\psi_{-\vec{P}}(\vec{k})}\\
& = \bra{\psi_{\vec{P}}(\vec{k})}I^{-1}T^{-1}STI\ket{\psi_{\vec{P}}(\vec{k})}\\
& = \bra{\psi_{\vec{P}}(\vec{k})}(-S)\ket{\psi_{\vec{P}}(\vec{k})}\\
& = \expval{-S}_{\vec{P},\vec{k}},
\end{split}
\end{equation}
which indicates that the spin orientation can be reversed by switching the ferroelectric polarization. This analysis is in fact confirmed by our calculated results of the spin textures shown in Fig. 6(b)-(c), where the fully reversal of the out-of-plane spin orientation is achieved under opposite in-plane ferroelectric polarization. \textcolor{red}{Such an interesting property indicates that an electrically controllable PSH in $MX$ monolayer can be realized, which is very useful for operation in the spintronic devices.}

Thus far, we have predicted that the PSH with large spin splitting is achieved in the $MX$ monolayer. In particular, GeTe monolayer is promising for spintronics since it has the largest strength of the spin splitting ($\alpha=1.67$ eV\AA) among the $MX$ monolayer. Because the PSH is achieved on the spin-split bands near the VBM [Fig. 3(a)], $p$-type doping for spintronics is expected to be realized. Moreover, by injection the hole doping into the valence band of the $MX$ monolayer, it is possible to map the formation and evolution of the PSH state using near-filled scanning Kerr microscopy\cite{Rudge}, which allow us to resolve the features down to tens-nm scale with sub-ns time revolution. Finally, the hole-doped $MX$ monolayer can also be applied to explore current-induced spin polarization known as a Edelstein effect\cite{Edelstein} and associated spin-orbit torque\cite{Gambardella}, indicating that the present system is promising for spintronic devices.

\section{CONCLUSION}
In summary, by using first-principles DFT calculations, supplemented with symmetry analyses, we investigated the effect of the SOC on the electronic structures of the $MX$ monolayer. We found that due to $C_{2v}$ point group symmetry in the $MX$ monolayer, the unidirectional out-of-plane spin configurations are preserved, inducing the PSH state that is similar to the [110] Dresselhauss model \cite{Bernevig} observed on the [110]-oriented semiconductor QW. Our first-principle calculations on various $MX$ ($M$: Sn, Ge; $X$: S, Se, Te) monolayers confirmed that this PSH is observed at near the VBM, supporting large spin splitting and small wavelength of the spin polarization. More importantly, we observed a reversible out-of-plane spin orientations under opposite in-plane ferroelectric polarization, indicating that an electrically controllable PSH in $MX$ monolayer can be realized, which is promising for spintronic devices. 

Recently, there are a number of other 2D materials that are predicted to maintain the in-plane ferroelectricity and the $C_{2v}$ symmetry of the crystals. Therefore, it opens a possibility to further explore the achievable PSH states in these materials. Among them are coming from the 2D elemental group‐V (As, Sb, and Bi) monolayer with the puckered lattice structure \cite{Xiao,BLiu}. Therefore, it is expected that our predictions will stimulate further theoretical and experimental efforts in the exploration of the PSH state in the 2D-based ferroelectric materials, broadening the range of the 2D materials for future spintronic applications.

\begin{acknowledgments}

The first author (M.A.U. Absor) would like to thanks Nanomaterial Reserach Institute, Kanazawa University, Japan, for providing financial support during his research visit. This work was partly supported by Grants-in-Aid on Scientific Research (Grant No. 16K04875) from the Japan Society for the Promotion of Science (JSPS) and a JSPS Grant-in-Aid for Scientific Research on Innovative Areas ”Discrete Geometric Analysis for Materials Design” (Grant No. 18H04481). Part of this research was supported by PDUPT Research Grant (2019) and BOPTN Research Grant (2019), Universitas Gadjah Mada, Indonesia.

\end{acknowledgments}

\bibliography{Reference1}

\begin{thebibliography}{58}%
\makeatletter
\providecommand \@ifxundefined [1]{%
 \@ifx{#1\undefined}
}%
\providecommand \@ifnum [1]{%
 \ifnum #1\expandafter \@firstoftwo
 \else \expandafter \@secondoftwo
 \fi
}%
\providecommand \@ifx [1]{%
 \ifx #1\expandafter \@firstoftwo
 \else \expandafter \@secondoftwo
 \fi
}%
\providecommand \natexlab [1]{#1}%
\providecommand \enquote  [1]{``#1''}%
\providecommand \bibnamefont  [1]{#1}%
\providecommand \bibfnamefont [1]{#1}%
\providecommand \citenamefont [1]{#1}%
\providecommand \href@noop [0]{\@secondoftwo}%
\providecommand \href [0]{\begingroup \@sanitize@url \@href}%
\providecommand \@href[1]{\@@startlink{#1}\@@href}%
\providecommand \@@href[1]{\endgroup#1\@@endlink}%
\providecommand \@sanitize@url [0]{\catcode `\\12\catcode `\$12\catcode
  `\&12\catcode `\#12\catcode `\^12\catcode `\_12\catcode `\%12\relax}%
\providecommand \@@startlink[1]{}%
\providecommand \@@endlink[0]{}%
\providecommand \url  [0]{\begingroup\@sanitize@url \@url }%
\providecommand \@url [1]{\endgroup\@href {#1}{\urlprefix }}%
\providecommand \urlprefix  [0]{URL }%
\providecommand \Eprint [0]{\href }%
\providecommand \doibase [0]{http://dx.doi.org/}%
\providecommand \selectlanguage [0]{\@gobble}%
\providecommand \bibinfo  [0]{\@secondoftwo}%
\providecommand \bibfield  [0]{\@secondoftwo}%
\providecommand \translation [1]{[#1]}%
\providecommand \BibitemOpen [0]{}%
\providecommand \bibitemStop [0]{}%
\providecommand \bibitemNoStop [0]{.\EOS\space}%
\providecommand \EOS [0]{\spacefactor3000\relax}%
\providecommand \BibitemShut  [1]{\csname bibitem#1\endcsname}%
\let\auto@bib@innerbib\@empty
\bibitem [{\citenamefont {Nitta}\ \emph {et~al.}(1997)\citenamefont {Nitta},
  \citenamefont {Akazaki}, \citenamefont {Takayanagi},\ and\ \citenamefont
  {Enoki}}]{Nitta}%
  \BibitemOpen
  \bibfield  {author} {\bibinfo {author} {\bibfnamefont {J.}~\bibnamefont
  {Nitta}}, \bibinfo {author} {\bibfnamefont {T.}~\bibnamefont {Akazaki}},
  \bibinfo {author} {\bibfnamefont {H.}~\bibnamefont {Takayanagi}}, \ and\
  \bibinfo {author} {\bibfnamefont {T.}~\bibnamefont {Enoki}},\ }\href
  {\doibase 10.1103/PhysRevLett.78.1335} {\bibfield  {journal} {\bibinfo
  {journal} {Phys. Rev. Lett.}\ }\textbf {\bibinfo {volume} {78}},\ \bibinfo
  {pages} {1335} (\bibinfo {year} {1997})}\BibitemShut {NoStop}%
\bibitem [{\citenamefont {Manchon}\ \emph {et~al.}(2015)\citenamefont
  {Manchon}, \citenamefont {Koo}, \citenamefont {Nitta}, \citenamefont
  {Frolov},\ and\ \citenamefont {Duine}}]{Manchon}%
  \BibitemOpen
  \bibfield  {author} {\bibinfo {author} {\bibfnamefont {A.}~\bibnamefont
  {Manchon}}, \bibinfo {author} {\bibfnamefont {H.~C.}\ \bibnamefont {Koo}},
  \bibinfo {author} {\bibfnamefont {J.}~\bibnamefont {Nitta}}, \bibinfo
  {author} {\bibfnamefont {S.~M.}\ \bibnamefont {Frolov}}, \ and\ \bibinfo
  {author} {\bibfnamefont {R.~A.}\ \bibnamefont {Duine}},\ }\href {\doibase
  10.1038/nmat4360} {\bibfield  {journal} {\bibinfo  {journal} {Nat. Matter}\
  }\textbf {\bibinfo {volume} {14}},\ \bibinfo {pages} {871} (\bibinfo {year}
  {2015})}\BibitemShut {NoStop}%
\bibitem [{\citenamefont {Dresselhaus}(1955)}]{Dresselhauss}%
  \BibitemOpen
  \bibfield  {author} {\bibinfo {author} {\bibfnamefont {G.}~\bibnamefont
  {Dresselhaus}},\ }\href {\doibase 10.1103/PhysRev.100.580} {\bibfield
  {journal} {\bibinfo  {journal} {Phys. Rev.}\ }\textbf {\bibinfo {volume}
  {100}},\ \bibinfo {pages} {580} (\bibinfo {year} {1955})}\BibitemShut
  {NoStop}%
\bibitem [{\citenamefont {Rashba}(1960)}]{Rashba}%
  \BibitemOpen
  \bibfield  {author} {\bibinfo {author} {\bibfnamefont {E.~I.}\ \bibnamefont
  {Rashba}},\ }\href@noop {} {\bibfield  {journal} {\bibinfo  {journal} {Sov.
  Phys. Solid State}\ }\textbf {\bibinfo {volume} {2}},\ \bibinfo {pages}
  {1224} (\bibinfo {year} {1960})}\BibitemShut {NoStop}%
\bibitem [{\citenamefont {Kuhlen}\ \emph {et~al.}(2012)\citenamefont {Kuhlen},
  \citenamefont {Schmalbuch}, \citenamefont {Hagedorn}, \citenamefont
  {Schlammes}, \citenamefont {Patt}, \citenamefont {Lepsa}, \citenamefont
  {G\"untherodt},\ and\ \citenamefont {Beschoten}}]{Kuhlen}%
  \BibitemOpen
  \bibfield  {author} {\bibinfo {author} {\bibfnamefont {S.}~\bibnamefont
  {Kuhlen}}, \bibinfo {author} {\bibfnamefont {K.}~\bibnamefont {Schmalbuch}},
  \bibinfo {author} {\bibfnamefont {M.}~\bibnamefont {Hagedorn}}, \bibinfo
  {author} {\bibfnamefont {P.}~\bibnamefont {Schlammes}}, \bibinfo {author}
  {\bibfnamefont {M.}~\bibnamefont {Patt}}, \bibinfo {author} {\bibfnamefont
  {M.}~\bibnamefont {Lepsa}}, \bibinfo {author} {\bibfnamefont
  {G.}~\bibnamefont {G\"untherodt}}, \ and\ \bibinfo {author} {\bibfnamefont
  {B.}~\bibnamefont {Beschoten}},\ }\href {\doibase
  10.1103/PhysRevLett.109.146603} {\bibfield  {journal} {\bibinfo  {journal}
  {Phys. Rev. Lett.}\ }\textbf {\bibinfo {volume} {109}},\ \bibinfo {pages}
  {146603} (\bibinfo {year} {2012})}\BibitemShut {NoStop}%
\bibitem [{\citenamefont {Qi}\ \emph {et~al.}(2006)\citenamefont {Qi},
  \citenamefont {Wu},\ and\ \citenamefont {Zhang}}]{Qi}%
  \BibitemOpen
  \bibfield  {author} {\bibinfo {author} {\bibfnamefont {X.-L.}\ \bibnamefont
  {Qi}}, \bibinfo {author} {\bibfnamefont {Y.-S.}\ \bibnamefont {Wu}}, \ and\
  \bibinfo {author} {\bibfnamefont {S.-C.}\ \bibnamefont {Zhang}},\ }\href
  {\doibase 10.1103/PhysRevB.74.085308} {\bibfield  {journal} {\bibinfo
  {journal} {Phys. Rev. B}\ }\textbf {\bibinfo {volume} {74}},\ \bibinfo
  {pages} {085308} (\bibinfo {year} {2006})}\BibitemShut {NoStop}%
\bibitem [{\citenamefont {Ganichev}\ \emph {et~al.}(2002)\citenamefont
  {Ganichev}, \citenamefont {Ivchenko}, \citenamefont {Bel'kov}, \citenamefont
  {Tarasenko}, \citenamefont {Sollinger}, \citenamefont {Weiss}, \citenamefont
  {Wegscheider},\ and\ \citenamefont {Prettl}}]{Ganichev}%
  \BibitemOpen
  \bibfield  {author} {\bibinfo {author} {\bibfnamefont {S.~D.}\ \bibnamefont
  {Ganichev}}, \bibinfo {author} {\bibfnamefont {E.~L.}\ \bibnamefont
  {Ivchenko}}, \bibinfo {author} {\bibfnamefont {V.~V.}\ \bibnamefont
  {Bel'kov}}, \bibinfo {author} {\bibfnamefont {S.~A.}\ \bibnamefont
  {Tarasenko}}, \bibinfo {author} {\bibfnamefont {M.}~\bibnamefont
  {Sollinger}}, \bibinfo {author} {\bibfnamefont {D.}~\bibnamefont {Weiss}},
  \bibinfo {author} {\bibfnamefont {W.}~\bibnamefont {Wegscheider}}, \ and\
  \bibinfo {author} {\bibfnamefont {W.}~\bibnamefont {Prettl}},\ }\href
  {\doibase 10.1038/417153a} {\bibfield  {journal} {\bibinfo  {journal}
  {Nature}\ }\textbf {\bibinfo {volume} {417}},\ \bibinfo {pages} {153}
  (\bibinfo {year} {2002})}\BibitemShut {NoStop}%
\bibitem [{\citenamefont {Lu}\ \emph {et~al.}(1998)\citenamefont {Lu},
  \citenamefont {Yau}, \citenamefont {Shukla}, \citenamefont {Shayegan},
  \citenamefont {Wissinger}, \citenamefont {R\"ossler},\ and\ \citenamefont
  {Winkler}}]{Lu}%
  \BibitemOpen
  \bibfield  {author} {\bibinfo {author} {\bibfnamefont {J.~P.}\ \bibnamefont
  {Lu}}, \bibinfo {author} {\bibfnamefont {J.~B.}\ \bibnamefont {Yau}},
  \bibinfo {author} {\bibfnamefont {S.~P.}\ \bibnamefont {Shukla}}, \bibinfo
  {author} {\bibfnamefont {M.}~\bibnamefont {Shayegan}}, \bibinfo {author}
  {\bibfnamefont {L.}~\bibnamefont {Wissinger}}, \bibinfo {author}
  {\bibfnamefont {U.}~\bibnamefont {R\"ossler}}, \ and\ \bibinfo {author}
  {\bibfnamefont {R.}~\bibnamefont {Winkler}},\ }\href {\doibase
  10.1103/PhysRevLett.81.1282} {\bibfield  {journal} {\bibinfo  {journal}
  {Phys. Rev. Lett.}\ }\textbf {\bibinfo {volume} {81}},\ \bibinfo {pages}
  {1282} (\bibinfo {year} {1998})}\BibitemShut {NoStop}%
\bibitem [{\citenamefont {Datta}\ and\ \citenamefont {Das}(1990)}]{Datta}%
  \BibitemOpen
  \bibfield  {author} {\bibinfo {author} {\bibfnamefont {S.}~\bibnamefont
  {Datta}}\ and\ \bibinfo {author} {\bibfnamefont {B.}~\bibnamefont {Das}},\
  }\href {\doibase http://dx.doi.org/10.1063/1.102730} {\bibfield  {journal}
  {\bibinfo  {journal} {Appl. Phys. Lett.}\ }\textbf {\bibinfo {volume} {56}},\
  \bibinfo {pages} {665} (\bibinfo {year} {1990})}\BibitemShut {NoStop}%
\bibitem [{\citenamefont {Chuang}\ \emph {et~al.}(2009)\citenamefont {Chuang},
  \citenamefont {Ho}, \citenamefont {Smith}, \citenamefont {Sfigakis},
  \citenamefont {Pepper}, \citenamefont {Chen}, \citenamefont {Fan},
  \citenamefont {Griffiths}, \citenamefont {Farrer}, \citenamefont {Beere},
  \citenamefont {Jones}, \citenamefont {Ritchie},\ and\ \citenamefont
  {Chen}}]{Chuang}%
  \BibitemOpen
  \bibfield  {author} {\bibinfo {author} {\bibfnamefont {P.}~\bibnamefont
  {Chuang}}, \bibinfo {author} {\bibfnamefont {S.-H.}\ \bibnamefont {Ho}},
  \bibinfo {author} {\bibfnamefont {L.~W.}\ \bibnamefont {Smith}}, \bibinfo
  {author} {\bibfnamefont {F.}~\bibnamefont {Sfigakis}}, \bibinfo {author}
  {\bibfnamefont {M.}~\bibnamefont {Pepper}}, \bibinfo {author} {\bibfnamefont
  {C.-H.}\ \bibnamefont {Chen}}, \bibinfo {author} {\bibfnamefont {J.-C.}\
  \bibnamefont {Fan}}, \bibinfo {author} {\bibfnamefont {J.~P.}\ \bibnamefont
  {Griffiths}}, \bibinfo {author} {\bibfnamefont {I.}~\bibnamefont {Farrer}},
  \bibinfo {author} {\bibfnamefont {H.~E.}\ \bibnamefont {Beere}}, \bibinfo
  {author} {\bibfnamefont {G.~A.~C.}\ \bibnamefont {Jones}}, \bibinfo {author}
  {\bibfnamefont {D.~A.}\ \bibnamefont {Ritchie}}, \ and\ \bibinfo {author}
  {\bibfnamefont {T.-M.}\ \bibnamefont {Chen}},\ }\href {\doibase
  10.1038/nnano.2014.296} {\bibfield  {journal} {\bibinfo  {journal} {Nature
  Nanotechnology}\ }\textbf {\bibinfo {volume} {10}},\ \bibinfo {pages} {35}
  (\bibinfo {year} {2009})}\BibitemShut {NoStop}%
\bibitem [{\citenamefont {Dyakonov}\ and\ \citenamefont
  {Perel}(1972)}]{Dyakonov}%
  \BibitemOpen
  \bibfield  {author} {\bibinfo {author} {\bibfnamefont {M.~I.}\ \bibnamefont
  {Dyakonov}}\ and\ \bibinfo {author} {\bibfnamefont {V.~I.}\ \bibnamefont
  {Perel}},\ }\href@noop {} {\bibfield  {journal} {\bibinfo  {journal} {Sov.
  Phys. Solid State}\ }\textbf {\bibinfo {volume} {13}},\ \bibinfo {pages}
  {3023} (\bibinfo {year} {1972})}\BibitemShut {NoStop}%
\bibitem [{\citenamefont {Bernevig}\ \emph {et~al.}(2006)\citenamefont
  {Bernevig}, \citenamefont {Orenstein},\ and\ \citenamefont
  {Zhang}}]{Bernevig}%
  \BibitemOpen
  \bibfield  {author} {\bibinfo {author} {\bibfnamefont {B.~A.}\ \bibnamefont
  {Bernevig}}, \bibinfo {author} {\bibfnamefont {J.}~\bibnamefont {Orenstein}},
  \ and\ \bibinfo {author} {\bibfnamefont {S.-C.}\ \bibnamefont {Zhang}},\
  }\href {\doibase 10.1103/PhysRevLett.97.236601} {\bibfield  {journal}
  {\bibinfo  {journal} {Phys. Rev. Lett.}\ }\textbf {\bibinfo {volume} {97}},\
  \bibinfo {pages} {236601} (\bibinfo {year} {2006})}\BibitemShut {NoStop}%
\bibitem [{\citenamefont {Schliemann}(2017)}]{Schliemann}%
  \BibitemOpen
  \bibfield  {author} {\bibinfo {author} {\bibfnamefont {J.}~\bibnamefont
  {Schliemann}},\ }\href {\doibase 10.1103/RevModPhys.89.011001} {\bibfield
  {journal} {\bibinfo  {journal} {Rev. Mod. Phys.}\ }\textbf {\bibinfo {volume}
  {89}},\ \bibinfo {pages} {011001} (\bibinfo {year} {2017})}\BibitemShut
  {NoStop}%
\bibitem [{\citenamefont {Altmann}\ \emph {et~al.}(2014)\citenamefont
  {Altmann}, \citenamefont {Walser}, \citenamefont {Reichl}, \citenamefont
  {Wegscheider},\ and\ \citenamefont {Salis}}]{Altmann}%
  \BibitemOpen
  \bibfield  {author} {\bibinfo {author} {\bibfnamefont {P.}~\bibnamefont
  {Altmann}}, \bibinfo {author} {\bibfnamefont {M.~P.}\ \bibnamefont {Walser}},
  \bibinfo {author} {\bibfnamefont {C.}~\bibnamefont {Reichl}}, \bibinfo
  {author} {\bibfnamefont {W.}~\bibnamefont {Wegscheider}}, \ and\ \bibinfo
  {author} {\bibfnamefont {G.}~\bibnamefont {Salis}},\ }\href {\doibase
  10.1103/PhysRevB.90.201306} {\bibfield  {journal} {\bibinfo  {journal} {Phys.
  Rev. B}\ }\textbf {\bibinfo {volume} {90}},\ \bibinfo {pages} {201306}
  (\bibinfo {year} {2014})}\BibitemShut {NoStop}%
\bibitem [{\citenamefont {Koralek}\ \emph {et~al.}(2009)\citenamefont
  {Koralek}, \citenamefont {Weber}, \citenamefont {Orenstein}, \citenamefont
  {Bernevig}, \citenamefont {Zhang}, \citenamefont {Mack},\ and\ \citenamefont
  {Awschalom}}]{Koralek}%
  \BibitemOpen
  \bibfield  {author} {\bibinfo {author} {\bibfnamefont {J.~D.}\ \bibnamefont
  {Koralek}}, \bibinfo {author} {\bibfnamefont {C.~P.}\ \bibnamefont {Weber}},
  \bibinfo {author} {\bibfnamefont {J.}~\bibnamefont {Orenstein}}, \bibinfo
  {author} {\bibfnamefont {B.~A.}\ \bibnamefont {Bernevig}}, \bibinfo {author}
  {\bibfnamefont {S.-C.}\ \bibnamefont {Zhang}}, \bibinfo {author}
  {\bibfnamefont {S.}~\bibnamefont {Mack}}, \ and\ \bibinfo {author}
  {\bibfnamefont {D.~D.}\ \bibnamefont {Awschalom}},\ }\href {\doibase
  10.1038/nature07871} {\bibfield  {journal} {\bibinfo  {journal} {Nature}\
  }\textbf {\bibinfo {volume} {458}},\ \bibinfo {pages} {610} (\bibinfo {year}
  {2009})}\BibitemShut {NoStop}%
\bibitem [{\citenamefont {Walser}\ \emph {et~al.}(2012)\citenamefont {Walser},
  \citenamefont {Reichl}, \citenamefont {egscheider},\ and\ \citenamefont
  {Salis}}]{Walser}%
  \BibitemOpen
  \bibfield  {author} {\bibinfo {author} {\bibfnamefont {M.~P.}\ \bibnamefont
  {Walser}}, \bibinfo {author} {\bibfnamefont {C.}~\bibnamefont {Reichl}},
  \bibinfo {author} {\bibfnamefont {W.}~\bibnamefont {egscheider}}, \ and\
  \bibinfo {author} {\bibfnamefont {G.}~\bibnamefont {Salis}},\ }\href
  {\doibase 10.1038/nphys2383} {\bibfield  {journal} {\bibinfo  {journal}
  {Nature Physics}\ }\textbf {\bibinfo {volume} {8}},\ \bibinfo {pages} {757}
  (\bibinfo {year} {2012})}\BibitemShut {NoStop}%
\bibitem [{\citenamefont {Sch\"onhuber}\ \emph {et~al.}(2014)\citenamefont
  {Sch\"onhuber}, \citenamefont {Walser}, \citenamefont {Salis}, \citenamefont
  {Reichl}, \citenamefont {Wegscheider}, \citenamefont {Korn},\ and\
  \citenamefont {Sch\"uller}}]{Schonhuber}%
  \BibitemOpen
  \bibfield  {author} {\bibinfo {author} {\bibfnamefont {C.}~\bibnamefont
  {Sch\"onhuber}}, \bibinfo {author} {\bibfnamefont {M.~P.}\ \bibnamefont
  {Walser}}, \bibinfo {author} {\bibfnamefont {G.}~\bibnamefont {Salis}},
  \bibinfo {author} {\bibfnamefont {C.}~\bibnamefont {Reichl}}, \bibinfo
  {author} {\bibfnamefont {W.}~\bibnamefont {Wegscheider}}, \bibinfo {author}
  {\bibfnamefont {T.}~\bibnamefont {Korn}}, \ and\ \bibinfo {author}
  {\bibfnamefont {C.}~\bibnamefont {Sch\"uller}},\ }\href {\doibase
  10.1103/PhysRevB.89.085406} {\bibfield  {journal} {\bibinfo  {journal} {Phys.
  Rev. B}\ }\textbf {\bibinfo {volume} {89}},\ \bibinfo {pages} {085406}
  (\bibinfo {year} {2014})}\BibitemShut {NoStop}%
\bibitem [{\citenamefont {Ishihara}\ \emph {et~al.}(2014)\citenamefont
  {Ishihara}, \citenamefont {Ohno},\ and\ \citenamefont {Ohno}}]{Ishihara}%
  \BibitemOpen
  \bibfield  {author} {\bibinfo {author} {\bibfnamefont {J.}~\bibnamefont
  {Ishihara}}, \bibinfo {author} {\bibfnamefont {Y.}~\bibnamefont {Ohno}}, \
  and\ \bibinfo {author} {\bibfnamefont {H.}~\bibnamefont {Ohno}},\ }\href
  {http://stacks.iop.org/1882-0786/7/i=1/a=013001} {\bibfield  {journal}
  {\bibinfo  {journal} {Applied Physics Express}\ }\textbf {\bibinfo {volume}
  {7}},\ \bibinfo {pages} {013001} (\bibinfo {year} {2014})}\BibitemShut
  {NoStop}%
\bibitem [{\citenamefont {Kohda}\ \emph {et~al.}(2012)\citenamefont {Kohda},
  \citenamefont {Lechner}, \citenamefont {Kunihashi}, \citenamefont
  {Dollinger}, \citenamefont {Olbrich}, \citenamefont {Sch\"onhuber},
  \citenamefont {Caspers}, \citenamefont {Bel'kov}, \citenamefont {Golub},
  \citenamefont {Weiss}, \citenamefont {Richter}, \citenamefont {Nitta},\ and\
  \citenamefont {Ganichev}}]{Kohda}%
  \BibitemOpen
  \bibfield  {author} {\bibinfo {author} {\bibfnamefont {M.}~\bibnamefont
  {Kohda}}, \bibinfo {author} {\bibfnamefont {V.}~\bibnamefont {Lechner}},
  \bibinfo {author} {\bibfnamefont {Y.}~\bibnamefont {Kunihashi}}, \bibinfo
  {author} {\bibfnamefont {T.}~\bibnamefont {Dollinger}}, \bibinfo {author}
  {\bibfnamefont {P.}~\bibnamefont {Olbrich}}, \bibinfo {author} {\bibfnamefont
  {C.}~\bibnamefont {Sch\"onhuber}}, \bibinfo {author} {\bibfnamefont
  {I.}~\bibnamefont {Caspers}}, \bibinfo {author} {\bibfnamefont {V.~V.}\
  \bibnamefont {Bel'kov}}, \bibinfo {author} {\bibfnamefont {L.~E.}\
  \bibnamefont {Golub}}, \bibinfo {author} {\bibfnamefont {D.}~\bibnamefont
  {Weiss}}, \bibinfo {author} {\bibfnamefont {K.}~\bibnamefont {Richter}},
  \bibinfo {author} {\bibfnamefont {J.}~\bibnamefont {Nitta}}, \ and\ \bibinfo
  {author} {\bibfnamefont {S.~D.}\ \bibnamefont {Ganichev}},\ }\href {\doibase
  10.1103/PhysRevB.86.081306} {\bibfield  {journal} {\bibinfo  {journal} {Phys.
  Rev. B}\ }\textbf {\bibinfo {volume} {86}},\ \bibinfo {pages} {081306}
  (\bibinfo {year} {2012})}\BibitemShut {NoStop}%
\bibitem [{\citenamefont {Sasaki}\ \emph {et~al.}(2014)\citenamefont {Sasaki},
  \citenamefont {Nonaka}, \citenamefont {Kunihashi}, \citenamefont {Kohda},
  \citenamefont {Bauernfeind}, \citenamefont {Dollinger}, \citenamefont
  {ARichter},\ and\ \citenamefont {Nitta}}]{Sasaki}%
  \BibitemOpen
  \bibfield  {author} {\bibinfo {author} {\bibfnamefont {A.}~\bibnamefont
  {Sasaki}}, \bibinfo {author} {\bibfnamefont {S.}~\bibnamefont {Nonaka}},
  \bibinfo {author} {\bibfnamefont {Y.}~\bibnamefont {Kunihashi}}, \bibinfo
  {author} {\bibfnamefont {M.}~\bibnamefont {Kohda}}, \bibinfo {author}
  {\bibfnamefont {T.}~\bibnamefont {Bauernfeind}}, \bibinfo {author}
  {\bibfnamefont {T.}~\bibnamefont {Dollinger}}, \bibinfo {author}
  {\bibfnamefont {K.}~\bibnamefont {ARichter}}, \ and\ \bibinfo {author}
  {\bibfnamefont {J.}~\bibnamefont {Nitta}},\ }\href {\doibase
  10.1038/nnano.2014.128} {\bibfield  {journal} {\bibinfo  {journal} {Nature
  Nanotechnology}\ }\textbf {\bibinfo {volume} {9}},\ \bibinfo {pages} {703}
  (\bibinfo {year} {2014})}\BibitemShut {NoStop}%
\bibitem [{\citenamefont {Chen}\ \emph {et~al.}(2014)\citenamefont {Chen},
  \citenamefont {F\"alt}, \citenamefont {Wegscheider},\ and\ \citenamefont
  {Salis}}]{Chen}%
  \BibitemOpen
  \bibfield  {author} {\bibinfo {author} {\bibfnamefont {Y.~S.}\ \bibnamefont
  {Chen}}, \bibinfo {author} {\bibfnamefont {S.}~\bibnamefont {F\"alt}},
  \bibinfo {author} {\bibfnamefont {W.}~\bibnamefont {Wegscheider}}, \ and\
  \bibinfo {author} {\bibfnamefont {G.}~\bibnamefont {Salis}},\ }\href
  {\doibase 10.1103/PhysRevB.90.121304} {\bibfield  {journal} {\bibinfo
  {journal} {Phys. Rev. B}\ }\textbf {\bibinfo {volume} {90}},\ \bibinfo
  {pages} {121304} (\bibinfo {year} {2014})}\BibitemShut {NoStop}%
\bibitem [{\citenamefont {Yamaguchi}\ and\ \citenamefont
  {Ishii}(2017)}]{Yamaguchi}%
  \BibitemOpen
  \bibfield  {author} {\bibinfo {author} {\bibfnamefont {N.}~\bibnamefont
  {Yamaguchi}}\ and\ \bibinfo {author} {\bibfnamefont {F.}~\bibnamefont
  {Ishii}},\ }\href {http://stacks.iop.org/1882-0786/10/i=12/a=123003}
  {\bibfield  {journal} {\bibinfo  {journal} {Applied Physics Express}\
  }\textbf {\bibinfo {volume} {10}},\ \bibinfo {pages} {123003} (\bibinfo
  {year} {2017})}\BibitemShut {NoStop}%
\bibitem [{\citenamefont {Absor}\ \emph {et~al.}(2015)\citenamefont {Absor},
  \citenamefont {Ishii}, \citenamefont {Kotaka},\ and\ \citenamefont
  {Saito}}]{Absor1}%
  \BibitemOpen
  \bibfield  {author} {\bibinfo {author} {\bibfnamefont {M.~A.~U.}\
  \bibnamefont {Absor}}, \bibinfo {author} {\bibfnamefont {F.}~\bibnamefont
  {Ishii}}, \bibinfo {author} {\bibfnamefont {H.}~\bibnamefont {Kotaka}}, \
  and\ \bibinfo {author} {\bibfnamefont {M.}~\bibnamefont {Saito}},\ }\href
  {http://stacks.iop.org/1882-0786/8/i=7/a=073006} {\bibfield  {journal}
  {\bibinfo  {journal} {Applied Physics Express}\ }\textbf {\bibinfo {volume}
  {8}},\ \bibinfo {pages} {073006} (\bibinfo {year} {2015})}\BibitemShut
  {NoStop}%
\bibitem [{\citenamefont {Absor}\ and\ \citenamefont {Ishii}(2019)}]{Absor2}%
  \BibitemOpen
  \bibfield  {author} {\bibinfo {author} {\bibfnamefont {M.~A.~U.}\
  \bibnamefont {Absor}}\ and\ \bibinfo {author} {\bibfnamefont
  {F.}~\bibnamefont {Ishii}},\ }\href {\doibase 10.1103/PhysRevB.99.075136}
  {\bibfield  {journal} {\bibinfo  {journal} {Phys. Rev. B}\ }\textbf {\bibinfo
  {volume} {99}},\ \bibinfo {pages} {075136} (\bibinfo {year}
  {2019})}\BibitemShut {NoStop}%
\bibitem [{\citenamefont {Ai}\ \emph {et~al.}(2019)\citenamefont {Ai},
  \citenamefont {Ma}, \citenamefont {Shao}, \citenamefont {Li},\ and\
  \citenamefont {Zhao}}]{Ai}%
  \BibitemOpen
  \bibfield  {author} {\bibinfo {author} {\bibfnamefont {H.}~\bibnamefont
  {Ai}}, \bibinfo {author} {\bibfnamefont {X.}~\bibnamefont {Ma}}, \bibinfo
  {author} {\bibfnamefont {X.}~\bibnamefont {Shao}}, \bibinfo {author}
  {\bibfnamefont {W.}~\bibnamefont {Li}}, \ and\ \bibinfo {author}
  {\bibfnamefont {M.}~\bibnamefont {Zhao}},\ }\href {\doibase
  10.1103/PhysRevMaterials.3.054407} {\bibfield  {journal} {\bibinfo  {journal}
  {Phys. Rev. Materials}\ }\textbf {\bibinfo {volume} {3}},\ \bibinfo {pages}
  {054407} (\bibinfo {year} {2019})}\BibitemShut {NoStop}%
\bibitem [{\citenamefont {Fei}\ \emph {et~al.}(2016)\citenamefont {Fei},
  \citenamefont {Kang},\ and\ \citenamefont {Yang}}]{Fei}%
  \BibitemOpen
  \bibfield  {author} {\bibinfo {author} {\bibfnamefont {R.}~\bibnamefont
  {Fei}}, \bibinfo {author} {\bibfnamefont {W.}~\bibnamefont {Kang}}, \ and\
  \bibinfo {author} {\bibfnamefont {L.}~\bibnamefont {Yang}},\ }\href {\doibase
  10.1103/PhysRevLett.117.097601} {\bibfield  {journal} {\bibinfo  {journal}
  {Phys. Rev. Lett.}\ }\textbf {\bibinfo {volume} {117}},\ \bibinfo {pages}
  {097601} (\bibinfo {year} {2016})}\BibitemShut {NoStop}%
\bibitem [{\citenamefont {Barraza-Lopez}\ \emph {et~al.}(2018)\citenamefont
  {Barraza-Lopez}, \citenamefont {Kaloni}, \citenamefont {Poudel},\ and\
  \citenamefont {Kumar}}]{Lopez}%
  \BibitemOpen
  \bibfield  {author} {\bibinfo {author} {\bibfnamefont {S.}~\bibnamefont
  {Barraza-Lopez}}, \bibinfo {author} {\bibfnamefont {T.~P.}\ \bibnamefont
  {Kaloni}}, \bibinfo {author} {\bibfnamefont {S.~P.}\ \bibnamefont {Poudel}},
  \ and\ \bibinfo {author} {\bibfnamefont {P.}~\bibnamefont {Kumar}},\ }\href
  {\doibase 10.1103/PhysRevB.97.024110} {\bibfield  {journal} {\bibinfo
  {journal} {Phys. Rev. B}\ }\textbf {\bibinfo {volume} {97}},\ \bibinfo
  {pages} {024110} (\bibinfo {year} {2018})}\BibitemShut {NoStop}%
\bibitem [{\citenamefont {Kaloni}\ \emph {et~al.}(2019)\citenamefont {Kaloni},
  \citenamefont {Chang}, \citenamefont {Miller}, \citenamefont {Xue},
  \citenamefont {Chen}, \citenamefont {Ji}, \citenamefont {Parkin},\ and\
  \citenamefont {Barraza-Lopez}}]{Kaloni}%
  \BibitemOpen
  \bibfield  {author} {\bibinfo {author} {\bibfnamefont {T.~P.}\ \bibnamefont
  {Kaloni}}, \bibinfo {author} {\bibfnamefont {K.}~\bibnamefont {Chang}},
  \bibinfo {author} {\bibfnamefont {B.~J.}\ \bibnamefont {Miller}}, \bibinfo
  {author} {\bibfnamefont {Q.-K.}\ \bibnamefont {Xue}}, \bibinfo {author}
  {\bibfnamefont {X.}~\bibnamefont {Chen}}, \bibinfo {author} {\bibfnamefont
  {S.-H.}\ \bibnamefont {Ji}}, \bibinfo {author} {\bibfnamefont {S.~S.~P.}\
  \bibnamefont {Parkin}}, \ and\ \bibinfo {author} {\bibfnamefont
  {S.}~\bibnamefont {Barraza-Lopez}},\ }\href {\doibase
  10.1103/PhysRevB.99.134108} {\bibfield  {journal} {\bibinfo  {journal} {Phys.
  Rev. B}\ }\textbf {\bibinfo {volume} {99}},\ \bibinfo {pages} {134108}
  (\bibinfo {year} {2019})}\BibitemShut {NoStop}%
\bibitem [{\citenamefont {Wan}\ \emph {et~al.}(2017)\citenamefont {Wan},
  \citenamefont {Liu}, \citenamefont {Xiao},\ and\ \citenamefont {Yao}}]{WWan}%
  \BibitemOpen
  \bibfield  {author} {\bibinfo {author} {\bibfnamefont {W.}~\bibnamefont
  {Wan}}, \bibinfo {author} {\bibfnamefont {C.}~\bibnamefont {Liu}}, \bibinfo
  {author} {\bibfnamefont {W.}~\bibnamefont {Xiao}}, \ and\ \bibinfo {author}
  {\bibfnamefont {Y.}~\bibnamefont {Yao}},\ }\href {\doibase 10.1063/1.4996171}
  {\bibfield  {journal} {\bibinfo  {journal} {Applied Physics Letters}\
  }\textbf {\bibinfo {volume} {111}},\ \bibinfo {pages} {132904} (\bibinfo
  {year} {2017})},\ \Eprint
  {http://arxiv.org/abs/https://doi.org/10.1063/1.4996171}
  {https://doi.org/10.1063/1.4996171} \BibitemShut {NoStop}%
\bibitem [{\citenamefont {Hanakata}\ \emph {et~al.}(2016)\citenamefont
  {Hanakata}, \citenamefont {Carvalho}, \citenamefont {Campbell},\ and\
  \citenamefont {Park}}]{Hanakata}%
  \BibitemOpen
  \bibfield  {author} {\bibinfo {author} {\bibfnamefont {P.~Z.}\ \bibnamefont
  {Hanakata}}, \bibinfo {author} {\bibfnamefont {A.}~\bibnamefont {Carvalho}},
  \bibinfo {author} {\bibfnamefont {D.~K.}\ \bibnamefont {Campbell}}, \ and\
  \bibinfo {author} {\bibfnamefont {H.~S.}\ \bibnamefont {Park}},\ }\href
  {\doibase 10.1103/PhysRevB.94.035304} {\bibfield  {journal} {\bibinfo
  {journal} {Phys. Rev. B}\ }\textbf {\bibinfo {volume} {94}},\ \bibinfo
  {pages} {035304} (\bibinfo {year} {2016})}\BibitemShut {NoStop}%
\bibitem [{\citenamefont {Perdew}\ \emph {et~al.}(1996)\citenamefont {Perdew},
  \citenamefont {Burke},\ and\ \citenamefont {Ernzerhof}}]{Perdew}%
  \BibitemOpen
  \bibfield  {author} {\bibinfo {author} {\bibfnamefont {J.~P.}\ \bibnamefont
  {Perdew}}, \bibinfo {author} {\bibfnamefont {K.}~\bibnamefont {Burke}}, \
  and\ \bibinfo {author} {\bibfnamefont {M.}~\bibnamefont {Ernzerhof}},\ }\href
  {\doibase 10.1103/PhysRevLett.77.3865} {\bibfield  {journal} {\bibinfo
  {journal} {Phys. Rev. Lett.}\ }\textbf {\bibinfo {volume} {77}},\ \bibinfo
  {pages} {3865} (\bibinfo {year} {1996})}\BibitemShut {NoStop}%
\bibitem [{\citenamefont {Ozaki}\ \emph {et~al.}(2009)\citenamefont {Ozaki},
  \citenamefont {Kino}, \citenamefont {Yu}, \citenamefont {Han}, \citenamefont
  {Kobayashi}, \citenamefont {Ohfuti}, \citenamefont {Ishii}, \citenamefont
  {Ohwaki}, \citenamefont {Weng},\ and\ \citenamefont {Terakura}}]{Openmx}%
  \BibitemOpen
  \bibfield  {author} {\bibinfo {author} {\bibfnamefont {T.}~\bibnamefont
  {Ozaki}}, \bibinfo {author} {\bibfnamefont {H.}~\bibnamefont {Kino}},
  \bibinfo {author} {\bibfnamefont {J.}~\bibnamefont {Yu}}, \bibinfo {author}
  {\bibfnamefont {M.~J.}\ \bibnamefont {Han}}, \bibinfo {author} {\bibfnamefont
  {N.}~\bibnamefont {Kobayashi}}, \bibinfo {author} {\bibfnamefont
  {M.}~\bibnamefont {Ohfuti}}, \bibinfo {author} {\bibfnamefont
  {F.}~\bibnamefont {Ishii}}, \bibinfo {author} {\bibfnamefont
  {T.}~\bibnamefont {Ohwaki}}, \bibinfo {author} {\bibfnamefont
  {H.}~\bibnamefont {Weng}}, \ and\ \bibinfo {author} {\bibfnamefont
  {K.}~\bibnamefont {Terakura}},\ }\href@noop {} {}\bibinfo {howpublished}
  {{http://www.openmx-square.org/}} (\bibinfo {year} {2009})\BibitemShut
  {NoStop}%
\bibitem [{\citenamefont {Troullier}\ and\ \citenamefont
  {Martins}(1991)}]{Troullier}%
  \BibitemOpen
  \bibfield  {author} {\bibinfo {author} {\bibfnamefont {N.}~\bibnamefont
  {Troullier}}\ and\ \bibinfo {author} {\bibfnamefont {J.~L.}\ \bibnamefont
  {Martins}},\ }\href {\doibase 10.1103/PhysRevB.43.1993} {\bibfield  {journal}
  {\bibinfo  {journal} {Phys. Rev. B}\ }\textbf {\bibinfo {volume} {43}},\
  \bibinfo {pages} {1993} (\bibinfo {year} {1991})}\BibitemShut {NoStop}%
\bibitem [{\citenamefont {Ozaki}(2003)}]{Ozaki}%
  \BibitemOpen
  \bibfield  {author} {\bibinfo {author} {\bibfnamefont {T.}~\bibnamefont
  {Ozaki}},\ }\href {\doibase 10.1103/PhysRevB.67.155108} {\bibfield  {journal}
  {\bibinfo  {journal} {Phys. Rev. B}\ }\textbf {\bibinfo {volume} {67}},\
  \bibinfo {pages} {155108} (\bibinfo {year} {2003})}\BibitemShut {NoStop}%
\bibitem [{\citenamefont {Ozaki}\ and\ \citenamefont {Kino}(2004)}]{Ozakikino}%
  \BibitemOpen
  \bibfield  {author} {\bibinfo {author} {\bibfnamefont {T.}~\bibnamefont
  {Ozaki}}\ and\ \bibinfo {author} {\bibfnamefont {H.}~\bibnamefont {Kino}},\
  }\href {\doibase 10.1103/PhysRevB.69.195113} {\bibfield  {journal} {\bibinfo
  {journal} {Phys. Rev. B}\ }\textbf {\bibinfo {volume} {69}},\ \bibinfo
  {pages} {195113} (\bibinfo {year} {2004})}\BibitemShut {NoStop}%
\bibitem [{\citenamefont {Theurich}\ and\ \citenamefont
  {Hill}(2001)}]{Theurich}%
  \BibitemOpen
  \bibfield  {author} {\bibinfo {author} {\bibfnamefont {G.}~\bibnamefont
  {Theurich}}\ and\ \bibinfo {author} {\bibfnamefont {N.~A.}\ \bibnamefont
  {Hill}},\ }\href {\doibase 10.1103/PhysRevB.64.073106} {\bibfield  {journal}
  {\bibinfo  {journal} {Phys. Rev. B}\ }\textbf {\bibinfo {volume} {64}},\
  \bibinfo {pages} {073106} (\bibinfo {year} {2001})}\BibitemShut {NoStop}%
\bibitem [{\citenamefont {Absor}\ \emph
  {et~al.}(2018{\natexlab{a}})\citenamefont {Absor}, \citenamefont {Santoso},
  \citenamefont {Harsojo}, \citenamefont {Abraha}, \citenamefont {Kotaka},
  \citenamefont {Ishii},\ and\ \citenamefont {Saito}}]{Absor3}%
  \BibitemOpen
  \bibfield  {author} {\bibinfo {author} {\bibfnamefont {M.~A.~U.}\
  \bibnamefont {Absor}}, \bibinfo {author} {\bibfnamefont {I.}~\bibnamefont
  {Santoso}}, \bibinfo {author} {\bibnamefont {Harsojo}}, \bibinfo {author}
  {\bibfnamefont {K.}~\bibnamefont {Abraha}}, \bibinfo {author} {\bibfnamefont
  {H.}~\bibnamefont {Kotaka}}, \bibinfo {author} {\bibfnamefont
  {F.}~\bibnamefont {Ishii}}, \ and\ \bibinfo {author} {\bibfnamefont
  {M.}~\bibnamefont {Saito}},\ }\href {\doibase 10.1103/PhysRevB.97.205138}
  {\bibfield  {journal} {\bibinfo  {journal} {Phys. Rev. B}\ }\textbf {\bibinfo
  {volume} {97}},\ \bibinfo {pages} {205138} (\bibinfo {year}
  {2018}{\natexlab{a}})}\BibitemShut {NoStop}%
\bibitem [{\citenamefont {Absor}\ \emph {et~al.}(2016)\citenamefont {Absor},
  \citenamefont {Kotaka}, \citenamefont {Ishii},\ and\ \citenamefont
  {Saito}}]{Absor4}%
  \BibitemOpen
  \bibfield  {author} {\bibinfo {author} {\bibfnamefont {M.~A.~U.}\
  \bibnamefont {Absor}}, \bibinfo {author} {\bibfnamefont {H.}~\bibnamefont
  {Kotaka}}, \bibinfo {author} {\bibfnamefont {F.}~\bibnamefont {Ishii}}, \
  and\ \bibinfo {author} {\bibfnamefont {M.}~\bibnamefont {Saito}},\ }\href
  {\doibase 10.1103/PhysRevB.94.115131} {\bibfield  {journal} {\bibinfo
  {journal} {Phys. Rev. B}\ }\textbf {\bibinfo {volume} {94}},\ \bibinfo
  {pages} {115131} (\bibinfo {year} {2016})}\BibitemShut {NoStop}%
\bibitem [{\citenamefont {Absor}\ \emph {et~al.}(2017)\citenamefont {Absor},
  \citenamefont {Santoso}, \citenamefont {Harsojo}, \citenamefont {Abraha},
  \citenamefont {Kotaka}, \citenamefont {Ishii},\ and\ \citenamefont
  {Saito}}]{Absor5}%
  \BibitemOpen
  \bibfield  {author} {\bibinfo {author} {\bibfnamefont {M.~A.~U.}\
  \bibnamefont {Absor}}, \bibinfo {author} {\bibfnamefont {I.}~\bibnamefont
  {Santoso}}, \bibinfo {author} {\bibnamefont {Harsojo}}, \bibinfo {author}
  {\bibfnamefont {K.}~\bibnamefont {Abraha}}, \bibinfo {author} {\bibfnamefont
  {H.}~\bibnamefont {Kotaka}}, \bibinfo {author} {\bibfnamefont
  {F.}~\bibnamefont {Ishii}}, \ and\ \bibinfo {author} {\bibfnamefont
  {M.}~\bibnamefont {Saito}},\ }\href {\doibase 10.1063/1.5008475} {\bibfield
  {journal} {\bibinfo  {journal} {Journal of Applied Physics}\ }\textbf
  {\bibinfo {volume} {122}},\ \bibinfo {pages} {153905} (\bibinfo {year}
  {2017})}\BibitemShut {NoStop}%
\bibitem [{\citenamefont {Absor}\ \emph
  {et~al.}(2018{\natexlab{b}})\citenamefont {Absor}, \citenamefont {Kotaka},
  \citenamefont {Ishii},\ and\ \citenamefont {Saito}}]{Absor6}%
  \BibitemOpen
  \bibfield  {author} {\bibinfo {author} {\bibfnamefont {M.~A.~U.}\
  \bibnamefont {Absor}}, \bibinfo {author} {\bibfnamefont {H.}~\bibnamefont
  {Kotaka}}, \bibinfo {author} {\bibfnamefont {F.}~\bibnamefont {Ishii}}, \
  and\ \bibinfo {author} {\bibfnamefont {M.}~\bibnamefont {Saito}},\ }\href
  {http://stacks.iop.org/1347-4065/57/i=4S/a=04FP01} {\bibfield  {journal}
  {\bibinfo  {journal} {Japanese Journal of Applied Physics}\ }\textbf
  {\bibinfo {volume} {57}},\ \bibinfo {pages} {04FP01} (\bibinfo {year}
  {2018}{\natexlab{b}})}\BibitemShut {NoStop}%
\bibitem [{\citenamefont {Gomes}\ and\ \citenamefont {Carvalho}(2015)}]{Gomes}%
  \BibitemOpen
  \bibfield  {author} {\bibinfo {author} {\bibfnamefont {L.~C.}\ \bibnamefont
  {Gomes}}\ and\ \bibinfo {author} {\bibfnamefont {A.}~\bibnamefont
  {Carvalho}},\ }\href {\doibase 10.1103/PhysRevB.92.085406} {\bibfield
  {journal} {\bibinfo  {journal} {Phys. Rev. B}\ }\textbf {\bibinfo {volume}
  {92}},\ \bibinfo {pages} {085406} (\bibinfo {year} {2015})}\BibitemShut
  {NoStop}%
\bibitem [{\citenamefont {Appelbaum}\ and\ \citenamefont
  {Li}(2016)}]{Appelbaum}%
  \BibitemOpen
  \bibfield  {author} {\bibinfo {author} {\bibfnamefont {I.}~\bibnamefont
  {Appelbaum}}\ and\ \bibinfo {author} {\bibfnamefont {P.}~\bibnamefont {Li}},\
  }\href {\doibase 10.1103/PhysRevB.94.155124} {\bibfield  {journal} {\bibinfo
  {journal} {Phys. Rev. B}\ }\textbf {\bibinfo {volume} {94}},\ \bibinfo
  {pages} {155124} (\bibinfo {year} {2016})}\BibitemShut {NoStop}%
\bibitem [{\citenamefont {Henkelman}\ and\ \citenamefont
  {Jónsson}(2000)}]{NEB}%
  \BibitemOpen
  \bibfield  {author} {\bibinfo {author} {\bibfnamefont {G.}~\bibnamefont
  {Henkelman}}\ and\ \bibinfo {author} {\bibfnamefont {H.}~\bibnamefont
  {Jónsson}},\ }\href {\doibase 10.1063/1.1323224} {\bibfield  {journal}
  {\bibinfo  {journal} {The Journal of Chemical Physics}\ }\textbf {\bibinfo
  {volume} {113}},\ \bibinfo {pages} {9978} (\bibinfo {year} {2000})},\ \Eprint
  {http://arxiv.org/abs/https://doi.org/10.1063/1.1323224}
  {https://doi.org/10.1063/1.1323224} \BibitemShut {NoStop}%
\bibitem [{\citenamefont {King-Smith}\ and\ \citenamefont
  {Vanderbilt}(1993)}]{BP}%
  \BibitemOpen
  \bibfield  {author} {\bibinfo {author} {\bibfnamefont {R.~D.}\ \bibnamefont
  {King-Smith}}\ and\ \bibinfo {author} {\bibfnamefont {D.}~\bibnamefont
  {Vanderbilt}},\ }\href {\doibase 10.1103/PhysRevB.47.1651} {\bibfield
  {journal} {\bibinfo  {journal} {Phys. Rev. B}\ }\textbf {\bibinfo {volume}
  {47}},\ \bibinfo {pages} {1651} (\bibinfo {year} {1993})}\BibitemShut
  {NoStop}%
\bibitem [{\citenamefont {Xu}\ \emph {et~al.}(2017)\citenamefont {Xu},
  \citenamefont {Yang}, \citenamefont {Wang},\ and\ \citenamefont
  {Feng}}]{LXu}%
  \BibitemOpen
  \bibfield  {author} {\bibinfo {author} {\bibfnamefont {L.}~\bibnamefont
  {Xu}}, \bibinfo {author} {\bibfnamefont {M.}~\bibnamefont {Yang}}, \bibinfo
  {author} {\bibfnamefont {S.~J.}\ \bibnamefont {Wang}}, \ and\ \bibinfo
  {author} {\bibfnamefont {Y.~P.}\ \bibnamefont {Feng}},\ }\href {\doibase
  10.1103/PhysRevB.95.235434} {\bibfield  {journal} {\bibinfo  {journal} {Phys.
  Rev. B}\ }\textbf {\bibinfo {volume} {95}},\ \bibinfo {pages} {235434}
  (\bibinfo {year} {2017})}\BibitemShut {NoStop}%
\bibitem [{\citenamefont {Yao}\ \emph {et~al.}(2017)\citenamefont {Yao},
  \citenamefont {Wang}, \citenamefont {Huang}, \citenamefont {Deng},
  \citenamefont {Yan}, \citenamefont {Zhang}, \citenamefont {Miyamoto},
  \citenamefont {Okuda}, \citenamefont {Li}, \citenamefont {Wang},
  \citenamefont {Gao}, \citenamefont {Liu}, \citenamefont {duan},\ and\
  \citenamefont {Zhou}}]{WYao}%
  \BibitemOpen
  \bibfield  {author} {\bibinfo {author} {\bibfnamefont {W.}~\bibnamefont
  {Yao}}, \bibinfo {author} {\bibfnamefont {E.}~\bibnamefont {Wang}}, \bibinfo
  {author} {\bibfnamefont {H.}~\bibnamefont {Huang}}, \bibinfo {author}
  {\bibfnamefont {K.}~\bibnamefont {Deng}}, \bibinfo {author} {\bibfnamefont
  {M.}~\bibnamefont {Yan}}, \bibinfo {author} {\bibfnamefont {K.}~\bibnamefont
  {Zhang}}, \bibinfo {author} {\bibfnamefont {K.}~\bibnamefont {Miyamoto}},
  \bibinfo {author} {\bibfnamefont {T.}~\bibnamefont {Okuda}}, \bibinfo
  {author} {\bibfnamefont {L.}~\bibnamefont {Li}}, \bibinfo {author}
  {\bibfnamefont {Y.}~\bibnamefont {Wang}}, \bibinfo {author} {\bibfnamefont
  {H.}~\bibnamefont {Gao}}, \bibinfo {author} {\bibfnamefont {C.}~\bibnamefont
  {Liu}}, \bibinfo {author} {\bibfnamefont {W.}~\bibnamefont {duan}}, \ and\
  \bibinfo {author} {\bibfnamefont {S.}~\bibnamefont {Zhou}},\ }\href {\doibase
  10.1038/ncomms14216} {\bibfield  {journal} {\bibinfo  {journal} {Nat.
  Commn.}\ }\textbf {\bibinfo {volume} {8}},\ \bibinfo {pages} {14216}
  (\bibinfo {year} {2017})}\BibitemShut {NoStop}%
\bibitem [{\citenamefont {Liu}\ \emph {et~al.}(2013)\citenamefont {Liu},
  \citenamefont {Guo},\ and\ \citenamefont {Freeman}}]{QLiu}%
  \BibitemOpen
  \bibfield  {author} {\bibinfo {author} {\bibfnamefont {Q.}~\bibnamefont
  {Liu}}, \bibinfo {author} {\bibfnamefont {Y.}~\bibnamefont {Guo}}, \ and\
  \bibinfo {author} {\bibfnamefont {A.~J.}\ \bibnamefont {Freeman}},\ }\href
  {\doibase 10.1021/nl4027346} {\bibfield  {journal} {\bibinfo  {journal} {Nano
  Letters}\ }\textbf {\bibinfo {volume} {13}},\ \bibinfo {pages} {5264}
  (\bibinfo {year} {2013})},\ \bibinfo {note} {pMID: 24127876}\BibitemShut
  {NoStop}%
\bibitem [{\citenamefont {Singh}\ and\ \citenamefont {Romero}(2017)}]{SSingh}%
  \BibitemOpen
  \bibfield  {author} {\bibinfo {author} {\bibfnamefont {S.}~\bibnamefont
  {Singh}}\ and\ \bibinfo {author} {\bibfnamefont {A.~H.}\ \bibnamefont
  {Romero}},\ }\href {\doibase 10.1103/PhysRevB.95.165444} {\bibfield
  {journal} {\bibinfo  {journal} {Phys. Rev. B}\ }\textbf {\bibinfo {volume}
  {95}},\ \bibinfo {pages} {165444} (\bibinfo {year} {2017})}\BibitemShut
  {NoStop}%
\bibitem [{\citenamefont {Tao}\ and\ \citenamefont {Tsymbal}(2018)}]{LLTao}%
  \BibitemOpen
  \bibfield  {author} {\bibinfo {author} {\bibfnamefont {L.~L.}\ \bibnamefont
  {Tao}}\ and\ \bibinfo {author} {\bibfnamefont {E.~Y.}\ \bibnamefont
  {Tsymbal}},\ }\href {\doibase 10.1038/s41467-018-05137-0} {\bibfield
  {journal} {\bibinfo  {journal} {Nature Communications}\ }\textbf {\bibinfo
  {volume} {9}},\ \bibinfo {pages} {2763} (\bibinfo {year} {2018})}\BibitemShut
  {NoStop}%
\bibitem [{\citenamefont {Fiori}\ \emph {et~al.}(2014)\citenamefont {Fiori},
  \citenamefont {Bonaccorso}, \citenamefont {Iannaccone}, \citenamefont
  {Palacios}, \citenamefont {Neumaier}, \citenamefont {Seabaugh}, \citenamefont
  {Banerjee},\ and\ \citenamefont {Colombo}}]{Fiori}%
  \BibitemOpen
  \bibfield  {author} {\bibinfo {author} {\bibfnamefont {G.}~\bibnamefont
  {Fiori}}, \bibinfo {author} {\bibfnamefont {F.}~\bibnamefont {Bonaccorso}},
  \bibinfo {author} {\bibfnamefont {G.}~\bibnamefont {Iannaccone}}, \bibinfo
  {author} {\bibfnamefont {T.}~\bibnamefont {Palacios}}, \bibinfo {author}
  {\bibfnamefont {D.}~\bibnamefont {Neumaier}}, \bibinfo {author}
  {\bibfnamefont {A.}~\bibnamefont {Seabaugh}}, \bibinfo {author}
  {\bibfnamefont {S.~K.}\ \bibnamefont {Banerjee}}, \ and\ \bibinfo {author}
  {\bibfnamefont {L.}~\bibnamefont {Colombo}},\ }\href {\doibase
  10.1038/nnano.2014.207} {\bibfield  {journal} {\bibinfo  {journal} {Nature
  Nanotechnology}\ }\textbf {\bibinfo {volume} {83}},\ \bibinfo {pages} {768}
  (\bibinfo {year} {2014})}\BibitemShut {NoStop}%
\bibitem [{\citenamefont {Haeni}\ \emph {et~al.}(2004)\citenamefont {Haeni},
  \citenamefont {Irvin}, \citenamefont {Chang}, \citenamefont {Uecker},
  \citenamefont {Reiche}, \citenamefont {Li}, \citenamefont {Choudhury},
  \citenamefont {Tian}, \citenamefont {Hawley}, \citenamefont {Craigo},
  \citenamefont {Tagantsev}, \citenamefont {Pan}, \citenamefont {Streiffer},
  \citenamefont {Chen}, \citenamefont {Kirchoefer}, \citenamefont {J.},\ and\
  \citenamefont {Schlom}}]{Ba}%
  \BibitemOpen
  \bibfield  {author} {\bibinfo {author} {\bibfnamefont {J.~H.}\ \bibnamefont
  {Haeni}}, \bibinfo {author} {\bibfnamefont {P.}~\bibnamefont {Irvin}},
  \bibinfo {author} {\bibfnamefont {W.}~\bibnamefont {Chang}}, \bibinfo
  {author} {\bibfnamefont {R.}~\bibnamefont {Uecker}}, \bibinfo {author}
  {\bibfnamefont {P.}~\bibnamefont {Reiche}}, \bibinfo {author} {\bibfnamefont
  {Y.~L.}\ \bibnamefont {Li}}, \bibinfo {author} {\bibfnamefont
  {S.}~\bibnamefont {Choudhury}}, \bibinfo {author} {\bibfnamefont
  {W.}~\bibnamefont {Tian}}, \bibinfo {author} {\bibfnamefont {M.~E.}\
  \bibnamefont {Hawley}}, \bibinfo {author} {\bibfnamefont {B.}~\bibnamefont
  {Craigo}}, \bibinfo {author} {\bibfnamefont {A.~K.}\ \bibnamefont
  {Tagantsev}}, \bibinfo {author} {\bibfnamefont {X.~Q.}\ \bibnamefont {Pan}},
  \bibinfo {author} {\bibfnamefont {S.~K.}\ \bibnamefont {Streiffer}}, \bibinfo
  {author} {\bibfnamefont {L.~Q.}\ \bibnamefont {Chen}}, \bibinfo {author}
  {\bibfnamefont {S.~W.}\ \bibnamefont {Kirchoefer}}, \bibinfo {author}
  {\bibfnamefont {L.}~\bibnamefont {J.}}, \ and\ \bibinfo {author}
  {\bibfnamefont {D.~G.}\ \bibnamefont {Schlom}},\ }\href {\doibase
  10.1038/nature02773} {\bibfield  {journal} {\bibinfo  {journal} {Nature}\
  }\textbf {\bibinfo {volume} {430}},\ \bibinfo {pages} {758} (\bibinfo {year}
  {2004})}\BibitemShut {NoStop}%
\bibitem [{\citenamefont {Di~Sante}\ \emph {et~al.}(2013)\citenamefont
  {Di~Sante}, \citenamefont {Barone}, \citenamefont {Bertacco},\ and\
  \citenamefont {Picozzi}}]{DSante}%
  \BibitemOpen
  \bibfield  {author} {\bibinfo {author} {\bibfnamefont {D.}~\bibnamefont
  {Di~Sante}}, \bibinfo {author} {\bibfnamefont {P.}~\bibnamefont {Barone}},
  \bibinfo {author} {\bibfnamefont {R.}~\bibnamefont {Bertacco}}, \ and\
  \bibinfo {author} {\bibfnamefont {S.}~\bibnamefont {Picozzi}},\ }\href
  {\doibase 10.1002/adma.201203199} {\bibfield  {journal} {\bibinfo  {journal}
  {Advanced Materials}\ }\textbf {\bibinfo {volume} {25}},\ \bibinfo {pages}
  {509} (\bibinfo {year} {2013})},\ \Eprint
  {http://arxiv.org/abs/https://onlinelibrary.wiley.com/doi/pdf/10.1002/adma.201203199}
  {https://onlinelibrary.wiley.com/doi/pdf/10.1002/adma.201203199} \BibitemShut
  {NoStop}%
\bibitem [{\citenamefont {Kim}\ \emph {et~al.}(2014)\citenamefont {Kim},
  \citenamefont {Im}, \citenamefont {Freeman}, \citenamefont {Ihm},\ and\
  \citenamefont {Jin}}]{MKim}%
  \BibitemOpen
  \bibfield  {author} {\bibinfo {author} {\bibfnamefont {M.}~\bibnamefont
  {Kim}}, \bibinfo {author} {\bibfnamefont {J.}~\bibnamefont {Im}}, \bibinfo
  {author} {\bibfnamefont {A.~J.}\ \bibnamefont {Freeman}}, \bibinfo {author}
  {\bibfnamefont {J.}~\bibnamefont {Ihm}}, \ and\ \bibinfo {author}
  {\bibfnamefont {H.}~\bibnamefont {Jin}},\ }\href {\doibase
  10.1073/pnas.1405780111} {\bibfield  {journal} {\bibinfo  {journal}
  {Proceedings of the National Academy of Sciences}\ }\textbf {\bibinfo
  {volume} {111}},\ \bibinfo {pages} {6900} (\bibinfo {year} {2014})},\ \Eprint
  {http://arxiv.org/abs/https://www.pnas.org/content/111/19/6900.full.pdf}
  {https://www.pnas.org/content/111/19/6900.full.pdf} \BibitemShut {NoStop}%
\bibitem [{\citenamefont {Rudge}\ \emph {et~al.}(2015)\citenamefont {Rudge},
  \citenamefont {Xu}, \citenamefont {Kolthammer}, \citenamefont {Hong},\ and\
  \citenamefont {Choi}}]{Rudge}%
  \BibitemOpen
  \bibfield  {author} {\bibinfo {author} {\bibfnamefont {J.}~\bibnamefont
  {Rudge}}, \bibinfo {author} {\bibfnamefont {H.}~\bibnamefont {Xu}}, \bibinfo
  {author} {\bibfnamefont {J.}~\bibnamefont {Kolthammer}}, \bibinfo {author}
  {\bibfnamefont {Y.~K.}\ \bibnamefont {Hong}}, \ and\ \bibinfo {author}
  {\bibfnamefont {B.~C.}\ \bibnamefont {Choi}},\ }\href {\doibase
  10.1063/1.4907712} {\bibfield  {journal} {\bibinfo  {journal} {Review of
  Scientific Instruments}\ }\textbf {\bibinfo {volume} {86}},\ \bibinfo {pages}
  {023703} (\bibinfo {year} {2015})},\ \Eprint
  {http://arxiv.org/abs/https://doi.org/10.1063/1.4907712}
  {https://doi.org/10.1063/1.4907712} \BibitemShut {NoStop}%
\bibitem [{\citenamefont {Edelstein}(1990)}]{Edelstein}%
  \BibitemOpen
  \bibfield  {author} {\bibinfo {author} {\bibfnamefont {V.}~\bibnamefont
  {Edelstein}},\ }\href {\doibase https://doi.org/10.1016/0038-1098(90)90963-C}
  {\bibfield  {journal} {\bibinfo  {journal} {Solid State Communications}\
  }\textbf {\bibinfo {volume} {73}},\ \bibinfo {pages} {233 } (\bibinfo {year}
  {1990})}\BibitemShut {NoStop}%
\bibitem [{\citenamefont {Gambardella}\ and\ \citenamefont
  {Miron}(2011)}]{Gambardella}%
  \BibitemOpen
  \bibfield  {author} {\bibinfo {author} {\bibfnamefont {P.}~\bibnamefont
  {Gambardella}}\ and\ \bibinfo {author} {\bibfnamefont {I.~M.}\ \bibnamefont
  {Miron}},\ }\href {\doibase 10.1098/rsta.2010.0336} {\bibfield  {journal}
  {\bibinfo  {journal} {Philosophical Transactions of the Royal Society A:
  Mathematical, Physical and Engineering Sciences}\ }\textbf {\bibinfo {volume}
  {369}},\ \bibinfo {pages} {3175} (\bibinfo {year} {2011})},\ \Eprint
  {http://arxiv.org/abs/https://royalsocietypublishing.org/doi/pdf/10.1098/rsta.2010.0336}
  {https://royalsocietypublishing.org/doi/pdf/10.1098/rsta.2010.0336}
  \BibitemShut {NoStop}%
\bibitem [{\citenamefont {Xiao}\ \emph {et~al.}(2018)\citenamefont {Xiao},
  \citenamefont {Wang}, \citenamefont {Yang}, \citenamefont {Lu}, \citenamefont
  {Feng},\ and\ \citenamefont {Zhang}}]{Xiao}%
  \BibitemOpen
  \bibfield  {author} {\bibinfo {author} {\bibfnamefont {C.}~\bibnamefont
  {Xiao}}, \bibinfo {author} {\bibfnamefont {F.}~\bibnamefont {Wang}}, \bibinfo
  {author} {\bibfnamefont {S.~A.}\ \bibnamefont {Yang}}, \bibinfo {author}
  {\bibfnamefont {Y.}~\bibnamefont {Lu}}, \bibinfo {author} {\bibfnamefont
  {Y.}~\bibnamefont {Feng}}, \ and\ \bibinfo {author} {\bibfnamefont
  {S.}~\bibnamefont {Zhang}},\ }\href {\doibase 10.1002/adfm.201707383}
  {\bibfield  {journal} {\bibinfo  {journal} {Advanced Functional Materials}\
  }\textbf {\bibinfo {volume} {28}},\ \bibinfo {pages} {1707383} (\bibinfo
  {year} {2018})},\ \Eprint
  {http://arxiv.org/abs/https://onlinelibrary.wiley.com/doi/pdf/10.1002/adfm.201707383}
  {https://onlinelibrary.wiley.com/doi/pdf/10.1002/adfm.201707383} \BibitemShut
  {NoStop}%
\bibitem [{\citenamefont {Liu}\ \emph {et~al.}(2019)\citenamefont {Liu},
  \citenamefont {Niu}, \citenamefont {Fu}, \citenamefont {Xi}, \citenamefont
  {Lei},\ and\ \citenamefont {Quhe}}]{BLiu}%
  \BibitemOpen
  \bibfield  {author} {\bibinfo {author} {\bibfnamefont {B.}~\bibnamefont
  {Liu}}, \bibinfo {author} {\bibfnamefont {M.}~\bibnamefont {Niu}}, \bibinfo
  {author} {\bibfnamefont {J.}~\bibnamefont {Fu}}, \bibinfo {author}
  {\bibfnamefont {Z.}~\bibnamefont {Xi}}, \bibinfo {author} {\bibfnamefont
  {M.}~\bibnamefont {Lei}}, \ and\ \bibinfo {author} {\bibfnamefont
  {R.}~\bibnamefont {Quhe}},\ }\href {\doibase
  10.1103/PhysRevMaterials.3.054002} {\bibfield  {journal} {\bibinfo  {journal}
  {Phys. Rev. Materials}\ }\textbf {\bibinfo {volume} {3}},\ \bibinfo {pages}
  {054002} (\bibinfo {year} {2019})}\BibitemShut {NoStop}%
\end{thebibliography}%


\end{document}